\documentclass[reprint,amsmath,amssymb,aps,pra]{revtex4-1}

\usepackage{graphicx}
\usepackage[dvipsnames]{xcolor}
\usepackage{amsmath,amssymb,amsthm}
\usepackage{mathrsfs}
\usepackage{blkarray}
\usepackage{url}

\newcommand{\ket}[1]{\ensuremath{\vert{#1}\rangle}}
\newcommand{\bra}[1]{\ensuremath{\langle{#1}\vert}}

\newcommand{\sfrac}[2]{\ensuremath{{\textstyle\frac{#1}{#2}}}}
\newcommand{\half}[0]{\sfrac{1}{2}}
\newcommand{\Tr}{\ensuremath{\mathrm{Tr}}}
\newcommand{\kett}[1]{\ensuremath{\vert{#1}\rangle\!\rangle}}
\newcommand{\bbra}[1]{\ensuremath{\langle\!\langle{#1}\vert}}
\newcommand{\bbrakett}[2]{\ensuremath{\langle\!\langle{#1}\vert{#2}\rangle\!\rangle}}

\newcommand{\sref}[1]{{Supplement Sec.~\ref{#1}}}

\begin{document}

\title{Large-scale Lindblad learning from time-series data}
\author{Ewout van den Berg,$^{1,*}$ Brad Mitchell,$^2$ Ken Xuan Wei,$^1$ Moein Malekakhlagh$^1$}
\affiliation{\vspace*{8pt}$^1$IBM Quantum, IBM T.J.~Watson Research Center, Yorktown Heights, NY 10598, USA\\
$^{2}$IBM Quantum, Almaden Research Center, San Jose, 95120, USA}
 \email{evandenberg@us.ibm.com}

\date{\today}
\maketitle 

{\bf Learned open-quantum-system models have broad applications in
  device characterization, device calibration, realistic noise model
  construction, and bounding observable estimates in quantum
  simulation. To this end, we develop a highly scalable Lindblad
  learning protocol assuming an effective time-independent Lindblad
  generator with local interactions. For a given operation that can be
  applied repeatedly on a quantum computer, the protocol learns the
  Lindbladian by forming a linear system of equations for the model
  parameters in terms of a set of observable values and their
  gradients. The required gradient information is obtained by fitting
  time-series data with sums of exponentially damped sinusoids and
  differentiating those curves. We develop a robust curve-fitting
  procedure that finds a parsimonious representation of the data up to
  shot noise. We demonstrate the approach by learning the Lindbladian
  for a full layer of gates on a 156-qubit superconducting quantum
  processor, providing the first learning experiment of this kind. We
  study the effects of state-preparation and measurement errors and
  limitations on the operations that can be learned. For improved
  performance under readout errors, we propose an optional fine-tuning
  strategy that improves the fit between the time-evolved model and
  the measured data.}

Open quantum systems can be described by the Lindblad master equation
$\dot{\rho} = \mathcal{L}(\rho)$~\cite{LIN1976a,GOR1976KSa}. When used
as a model, the Lindbladian provides a principled physical description
of quantum operations implemented on a quantum processor and has
wide-ranging applications. For a start, these models can be used to
characterize and calibrate various operations for quantum
computations~\cite{SAM2022GBCa,VAR2025MBa,SCH2024SLa}. In the context
of error mitigation~\cite{TEM2017BGa, LI2017Ba, CAI2023BBEa}, learned
Lindbladians can be used to isolate noise affecting gates in quantum
circuits~\cite{BLU2022SNPa,MAL2025SPGa}, which can subsequently be
mitigated using quasi-probabilistic techniques. Finally, a more recent
and promising direction is the use of learned models to rigorously
bound errors in Hamiltonian simulation~\cite{PAS2022OKZa,KRA2025JLOa}.

For models to be useful, they first need to be learned. The two main
considerations for learning are accuracy of the learned model, and the
scalability of the learning protocol. Techniques such as quantum
process tomography (QPT)~\cite{POY1997CZa, CHU1997Na, NIE2010Ca,
  DAR2001La, MER2013GSPa,
  GRE2015a,GEB2023SGGa,OBR2004PGJa,SUR2022KKGa,AHM2023QKa} are well
established, but even with optimized ancilla-based measurement
schemes~\cite{LEU2000a,DAR2001La,ALT2003BJWa,MOH2006La}, their cost
scales exponentially in the number of qubits, thereby limiting their
use to only very small systems. Some recent tomographic learning
techniques improve scalability by leveraging locality or other
structure, e.g.,~\cite{BRI2023RKa,MIL2026OHSa}.

In this work, we develop a scalable protocol for learning the
time-independent Lindbladian of a given operator
$\Lambda = \exp(\tau\mathcal{L})$.  For this, we assume that the
Hamiltonian and dissipative terms can be expressed in terms of one-
and two-local Pauli operators $P_{\ell}$, such that~\cite{WOL2008Ca}:
\begin{equation}\label{Eq:GKSL}
\mathcal{L}(\rho) = -\frac{i}{\hbar} [H,\rho] + \sum_{ij} \beta_{ij}\left(P_i\rho
P_j^{\dag} - \half\{P_j^{\dag}P_i,\rho\}\right),
\end{equation}
with Hamiltonian $H = \sum_k \alpha_k P_k$. Although the protocol can
easily be extended to support additional (non-local) terms, doing so
does come at the cost of reduced scalability. Given $\Lambda$ and the
selected model terms, our protocol learns the model coefficients
$\alpha_k$ and $\beta_{ij}$. We assume access only to $\Lambda$ with
unit evolution time $\tau$ and exclude evolution of $\mathcal{L}$ with
fractional times. Omitting dissipative terms reduces the problem to
Hamiltonian learning.

There exists a variety of protocols for Lindblad learning, including
algorithms based on matrix logarithms and explicit time evolution
(see~\sref{SI:Background} for details and references). Both classes of
algorithms require the formation of matrices or vectors that grow
exponentially in the number of qubits, regardless of any locality
assumptions. A third class of
algorithms~\cite{ZUB2021YLBa,STI2024MDWa,OLS2024KKKa,AGU2025WDRa,BAI2020GPLa}
leverages Ehrenfest's theorem for the evolution of observable
expectation values:
\begin{equation}\label{Eq:Ehrenfest}
\frac{d}{dt}\langle O\rangle_t
=\langle O\mathcal{L}(\rho(t))\rangle.
\end{equation}
This approach enables more scalable learning algorithms and lays the
foundation for learning that is robust to state-preparation errors
since the equation is consistent for any initial
state~\cite{OLS2024KKKa}.

For a given observable $O$ we can expand the right-hand side
  of Eq.~\eqref{Eq:Ehrenfest} by substituting Eq.~\eqref{Eq:GKSL} and
  observing that the expectation is a linear operator:
\begin{align}\label{Eq:LinearExpression}
\begin{split}
\frac{d}{dt}\langle O\rangle_t
& =  \sum_j \alpha_j \langle -i[O,P_j]\rangle_t +\\
& \phantom{=}\ \sum_{ij}\sfrac{\beta_{ij}}{2}(
\langle P_j^{\dag}[O,P_i]\rangle + \langle
[P_j^{\dag},O]P_i\rangle_t).
\end{split}
\end{align}
That is, the gradient of the observable value is expressed as a linear
combination of the model parameters with coefficients given by the
expectation values $\langle O'\rangle$ for various observables $O'$
formed as the product of $O$ and the Pauli terms appearing in the
model (see~\sref{SI:Background} for details).

By applying Eq.~\eqref{Eq:LinearExpression} for different initial
states, observables, and evolution times we can form a linear system
of equations of the form $Ax=b$, where $x = [\alpha; \beta]$
represents the vectorized model parameters, the entries in matrix $A$
are given by the various observable expectation values appearing in
the right-hand side of Eq.~\ref{Eq:LinearExpression}, and $b$
represents a vector of different observable-value derivatives
$\sfrac{d}{dt}\langle O\rangle_t$.

For simple initial states $\rho(0)$ that can be prepared with high
fidelity, it is possible to classically evaluate the expectation
values in the right-hand side of Eq.~\eqref{Eq:LinearExpression} at
$t=0$, as done in~\cite{ZUB2021YLBa,STI2024MDWa}. For time-evolved
states $\rho(t)$, with $t > 0$, this approach is no longer possible
since classical time evolution would require access to the very
Lindbladian we are trying to learn. However, the observable values can
still be measured experimentally, as done in~\cite{AGU2025WDRa}.

The main challenge in Ehrenfest-based Lindblad learning is the
estimation of the derivatives on the left-hand side of
Eq.~\eqref{Eq:LinearExpression}. One option is to use finite
differencing~\cite{ZUB2021YLBa,AGU2025WDRa}, but this evaluates the
observables at $t$ and $t\pm \Delta t$ for some small $\Delta t$,
which requires access to fractional applications of the operator of
interest. Stilck Fran\c{c}a {\it{et al.}}~\cite{STI2024MDWa} propose
measuring observable values $\langle{O}\rangle_t$ at a sequence of
time points, followed by (robust) polynomial
fitting~\cite{KAN2017KPa}, which then enables estimation of the
gradient based on the fitted curves. In their method, the gradient is
evaluated only at $t=0$, which enables purely classical evaluation of
the matrix $A$ (assuming state-preparation errors are known or
absent).

\begin{figure*}[th]
\centering
\includegraphics[width=0.9\textwidth]{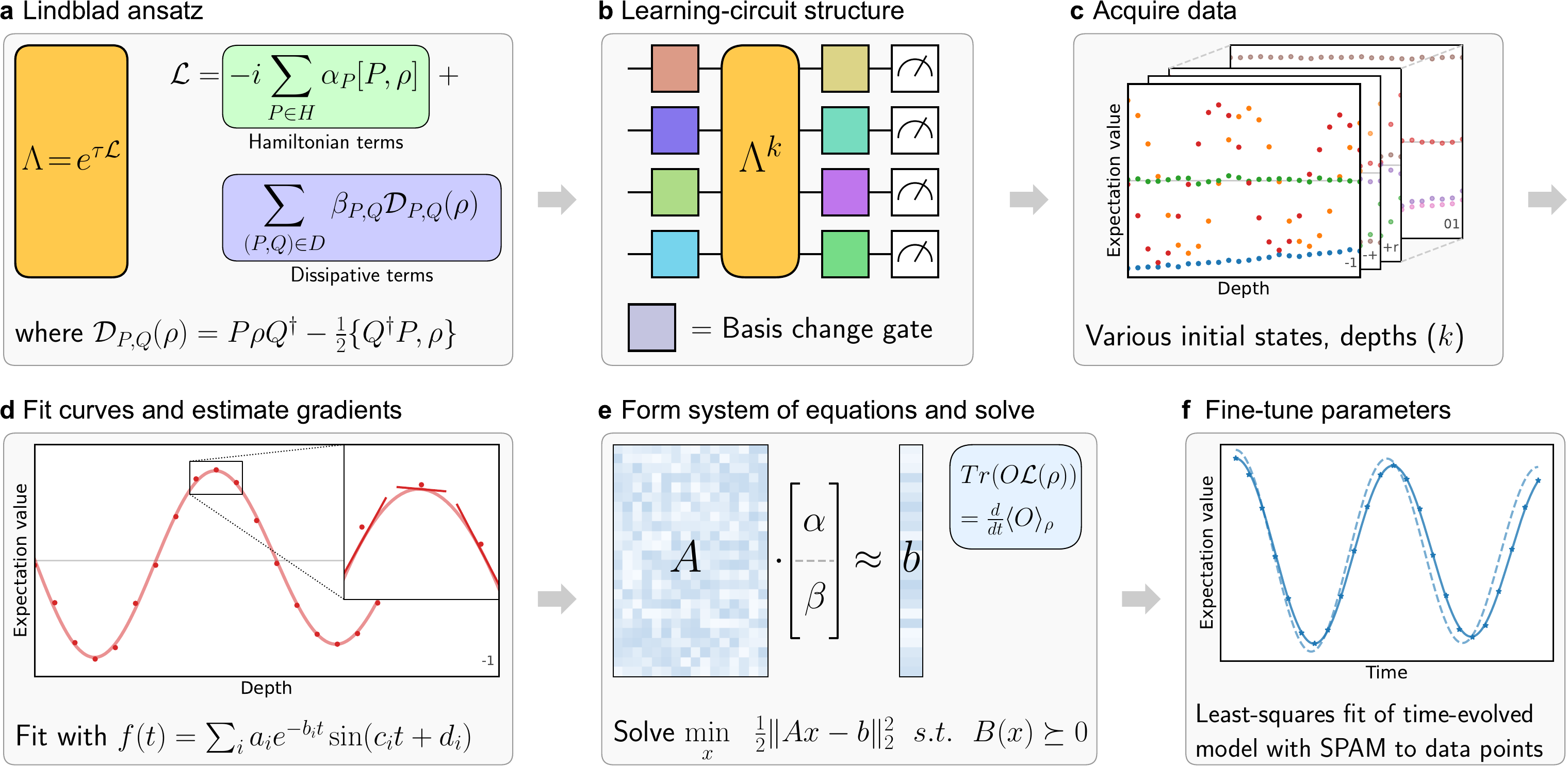}
\caption{Overview of the learning protocol: given
(a) an operation that we want to learn:
$\Lambda=\exp(\tau\mathcal{L})$, with unit evolution time $\tau$, and
Lindbladian $\mathcal{L}$ consisting of Hamiltonian and dissipative
terms, we
(b) prepare circuits that apply the operation an integer number of
times (depth $k$) flanked by single-qubit gates to implement
appropriate state-preparation and measurement basis changes as well as
readout twirling; we then
(c) select a series of (randomly selected) initial states. Per initial
state, select a set of Pauli observables $\{O\}$ and determine the set
$\{O'\}$ of associated observables given by the $[O,P_j]$,
$P_j^{\dag}[O,P_i]$, and $[P_j^{\dag},O]P_i$ terms in
Eq.~\eqref{Eq:LinearExpression}. Measure the observable expectation
value for each observable in $\{O\}\cup \{O'\}$ at a range of integer
depths $[0,k_{\max}]$;
(d) the observable expectation values at the different depths for a
single initial state and observable in $\{O\}$ define a time
series. In this step we fit each time series with a sum of
exponentially decaying sinusoids. (e) using the data points from step
(c) and curve fits from step (d), form a system of linear equations
$Ax=b$, where $x$ is a vector of unknown model parameters $\alpha$ and
$\beta$. Each combination of initial state, observable $O$, and depth
forms a single equation. The entries in $A$ are given by the
observable expectation values of the $O'$ terms generated using $O$,
as described above.  The corresponding entry in $b$ is obtained by
evaluating the derivative of the curve fit at the given depth. Once
the system of equations has been formed, we obtain the model
parameters by means of least-squares optimization, possibly including
a positive-semidefinite constraint $B(x)\succeq 0$, where $B$ is an
operator that extracts the $\beta$ terms from vector $x$ and reshapes
them into a matrix. An optional (f) fine-tuning stage can be added to
obtain model parameters (solid) that improve the fit between the
time-evolved initial model (dashed) and measurement data (dots). This
is done by including parameters for state-preparation and measurement
errors and minimizing a nonlinear least-squares objective using a
gradient-descent based method.  The fine-tuning stage requires
numerical time evolution of the model, which scales exponentially in
the number of qubits. As a result, fine tuning can only be applied to
small qubit patches in practice.}\label{Fig:Overview}
\end{figure*}

\paragraph*{Proposed learning protocol.}

\begin{figure*}[t!]
\centering
\includegraphics[width=0.98\textwidth]{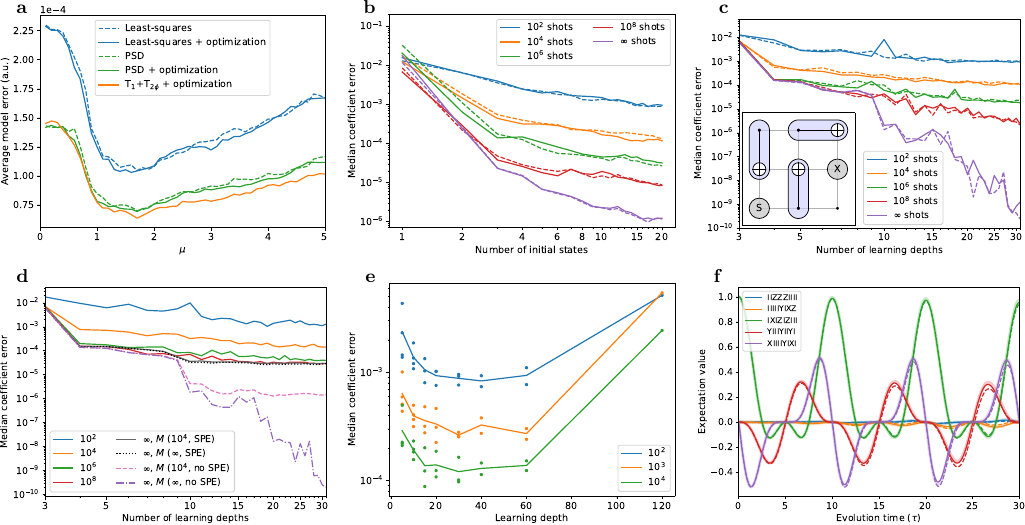}
\caption{Results for the 3$\times$3 circuit in inset (c) simulated
  without (a--c) and with (d--f) state-preparation and measurement
  (SPAM) errors. (a) The average model error as a function of misfit
  multiplier parameter $\mu$ in curve fitting for different
  optimization algorithms and models with 5 initial states and
  learning depth 30. The optimization suffix indicates that the fitted
  curves are locally optimized to better fit the data. The average
  error is taken as the geometric mean of the median absolute
  coefficient errors for setting with different shot counts. The
  median model coefficient error (b) as a function of the number of
  initial states, for learning depth 10 with semidefinite
  optimization, and (c) as a function of the learning depth for 10
  initial states. Similar results are obtained when learning with
  $2\times$ the nominal state-preparation error but no readout error
  (dashed lines). The peak in the blue curve when using ten learning
  depths is due to severe overfitting of one of the curves during fine
  tuning of the curve fit. Plot (d) shows the median coefficient error
  as a function of the number of learning depths (the maximum learning
  depth plus one) for simulated data points with SPAM
  errors. Readout-error mitigation requires estimation of the
  confusion matrix $M$, which is done using $10^4$ shots and
  state-preparation errors ($10^4$, SPE) unless otherwise
  indicated. Learning of $M$ is done both with and without
  state-preparation error (SPE). (e) The median model coefficient
  error for different combinations of learning depths, numbers of
  initial states, and shot count multipliers such that total shot
  count is kept fixed. The curves represent the average performance
  over all settings with a given learning depth for different base
  shot counts. (f) Expectation values for several weight-three Pauli
  observables for the initial state
  $\vert$-\hspace{1pt}-+1000\hspace{1pt}-\hspace{1pt}-$\rangle$
  time-evolved according to the ideal Lindbladian (thick faint lines)
  and the learned model based on 20 initial states, a maximum learning
  depth of 30, and $10^4$ shots both with (dashed) and without SPAM
  errors (solid). These observables are not used when learning the
  model.}\label{Fig:Simulations}
\end{figure*}

We develop a learning protocol that uses curve fitting for gradient
evaluation, but also leverages observable and gradient information at
depths other than $t=0$. Our learning protocol is summarized in
Fig.~\ref{Fig:Overview}.  First, we specify a Lindblad ansatz and the
Hamiltonian and dissipative terms to be learned (see
Fig.~\ref{Fig:Overview}(a)). Second, we construct circuits that
prepare selected initial states, repeatedly apply the target
operation, and measure the required Pauli observables at a range of
integer depths (see Fig.~\ref{Fig:Overview}(b,c)). Third, we fit each
observable time series with exponentially damped sinusoids and
differentiate the fitted curves to estimate the observable gradients,
as shown in Fig.~\ref{Fig:Overview}(d). Fourth, we combine the
measured expectation values and estimated gradients to form a linear
system for the Lindblad coefficients, which we solve by constrained or
unconstrained least squares (see Fig.~\ref{Fig:Overview}(e)). Finally,
for sufficiently small systems, the resulting model can optionally be
fine-tuned jointly with SPAM parameters by matching its time-evolved
predictions directly to the measured data, as illustrated in
Fig.~\ref{Fig:Overview}(f).

In particular, for a given initial state, which can be selected as
random Pauli eigenstates, and observable $O$, we measure
$\langle O\rangle_t$ at different evolution times $t$. We then fit
these data points with a suitable function $f(t)$ that allows us to
estimate $\sfrac{d}{dt}\langle O\rangle_t \approx f'(t)$ for
$t \geq 0$. When $O$ is a Pauli observable, the terms $[O, P_j]$,
$P_j^{\dag}[O,P_i]$, and $[P_j^{\dag},O]P_i$ in the right-hand side of
Eq.~\eqref{Eq:LinearExpression} give rise to new observables $O'$ that
are either scaled Pauli operators, or zero. We can experimentally
measure the expectation value of all non-zero observables $O'$ at the
various evaluation times. Substitution of the expectation values and
gradient estimate into Eq.~\eqref{Eq:LinearExpression} then gives a
linear relationship in terms of the $\alpha$ and $\beta$
terms. Repeating this for various initial states and observables
results in a system of equations. Unlike~\cite{STI2024MDWa}, the
system of equations includes gradient and observable values for
$t > 0$. Note that the entries in $A$ cannot be classically evaluated
for $t > 0$, since that would (i) require time evolution with the
Lindbladian we are trying to learn and (ii) have exponential
complexity in the number of qubits.

A single $O$ can result in a number of Pauli operators $O'$ that need
to be measured experimentally. The total number of unique observable
values that needs to be measured depends on the Pauli terms included
in the Lindblad model. For the method to be practical, we need to
limit the number of unique observables, and ideally choose them in a
way that allows us to measure them in as few measurement bases as
possible. We therefore choose the observables $O$ as one-local Pauli
terms, and assume that the Hamiltonian contains one- and two-local
Pauli terms following the qubit topology. For the dissipative part we
assume one-local Pauli terms with $\beta_{ij} = 0$ whenever the
support of $P_i$ and $P_j$ differs. As a result, the support for
observables $O'$ generated by dissipative terms remains one local. The
support of the $O'$ terms corresponding to Hamiltonian terms never
exceeds that of the corresponding $P_j$ (otherwise $O$ and $P_j$ would
commute), and therefore remains at most two local. For four-colorable
$n$-qubit topologies, we can measure all $3n$ one-local observables
$O$ as well as all corresponding $O'$, using nine carefully
constructed measurement bases~\cite{BER2024Wa}. That is, each
two-local Pauli operator $\{X,Y,Z\}^{\otimes 2}$ is included in at
least one basis, and can therefore be measured. Moreover, the same set
of measurement bases can be used to measure the expectation values for
all $3n$ one-local Pauli observables $O$ and their corresponding $O'$
terms. Extending the model beyond two-local terms results in
additional observables that need to be measured.

We show in~\sref{SI:ObservableValues} that observable expectation
values $\langle O\rangle_t$ are described by sums of exponentially
damped sinusoids, even when affected by (twirled) readout
errors~\cite{BER2022MTa}. We therefore fit the measured expectation
values by sums of exponentially damped sinusoids instead of low-degree
polynomials (see~\sref{SI:PolynomialFits}). For this, we develop a
specialized fitting protocol based on the generalized
pencil-of-function method~\cite{HUA1989Sa,SAR1995Pa}. To avoid
overfitting, we estimate the expected level of shot noise and find the
most parsimonious fit that approximates the data such that the misfit
is within a selected scalar multiple $\mu$ of the expected misfit (see
also~\sref{SI:CurveFitting} and~\ref{SI:ExpectedMisfit}).

Once we have constructed the system of equations, we obtain the model
parameters $x=[\alpha; \beta]$ by solving
\begin{equation}\label{Eq:SystemAx=b}
\mathop{\mbox{minimize}}_{x}\quad \half\Vert Ax-b\Vert_2^2\quad
\mbox{subject to}\quad B(x) \succeq 0
\end{equation}
using a splitting conic solver~\cite{ODO2016CPBa,ODO2021a}. The
positive-semidefinite constraint on the matrix representation $B(x)$
of the $\beta$ parameters ensures that the resulting Lindblad model is
physical. The learning protocol is highly scalable: the number of
model parameters scales linearly in the number of qubits when each
qubit is connected to a fixed number of other qubits. In case of
all-to-all qubit connectivity, the number scales quadratically. An
outline of overall learning procedure is summarized in
Figure~\ref{Fig:Overview}.

To better reflect the quantum circuit model, we assume access only to
the (noisy) $\Lambda$ operator, and exclude fractional evolution times
of $\mathcal{L}$. This restriction does limit the class of
Lindbladians that can be learned successfully.  In particular, we
cannot expect to recover Hamiltonian terms with coefficients above the
threshold set by the Nyquist theorem. Consider for instance the case
of learning $R_z(\theta) = \exp(-iH)$ with Hamiltonian
$H = \frac{\theta}{2}Z$.  With an initial state $\ket{0}$ and
evolution depth $k$, we have $\langle X\rangle = \cos(\theta k)$. To
resolve this function using curve fitting, we must sample it at least
at the Nyquist rate. Given that we can only sample at integer times
$k$, this rate is satisfied only when $\vert \theta\vert < \pi$. For
larger $\theta$ values, aliasing will occur and result in incorrect
curve fits and, therefore, inaccurate gradient estimates
(see~\sref{SI:Nyquist} for more details).

\paragraph*{Simulations}

Evaluation of the accuracy of the learning algorithm based on
experimental results is complicated by the fact that we do not know
the ground truth. We therefore run simulations on an imaginary
superconducting quantum processor with a $3\!\times\! 3$ grid of
qubits (labeled 0--8 from left to right, top to bottom). Per qubit, we
randomly sample $T_1$ and $T_{2\phi}$ times between
$100$--$200$$\mu{}$s and $50$--$150$$\mu{}$s, respectively, and use a
unit gate duration of 50ns. We assume single-qubit $Z$ errors at
5--20kHz rates and two-qubit $ZZ$ interactions between neighboring
qubits at 50--100kHz. The nominal state-preparation and measurement
(SPAM) error per qubit is taken between 0--1\% and 0--2\%,
respectively (for details, see~\sref{SI:SyntheticSetup}).  For the
simulations we define a Hamiltonian that generates CX gates on qubits
(0,3), (1,2), and (7,4), a phase gate on qubit 6, and an X gate on
qubit 5. We set the unit evolution time to 20\% of the unit gate time
to avoid aliasing effects. Time evolution is done by means of
Trotter-Suzuki product formulas~\cite{SUZ1990a,SUZ1991a}, which
approximate the time evolution as a series of short-term time
evolutions that are easier to evaluate. To ensure that errors due to
approximate time evolution do not dominate the final learning error,
we use a sixth-order Trotter-Suzuki product
formula~\cite{SUZ1990a,SUZ1991a} with 100 steps per unit evolution
time (see also~\sref{SI:SimulationImpact2x3}).

For the initial set of simulations we exclude SPAM errors. Since
accurate curve fitting is crucial to the overall learning performance,
we first consider the learning accuracy as a function of the misfit
multiplier $\mu$ in Fig.~\ref{Fig:Simulations}a.  Based on further
simulations in~\sref{SI:ChoiceMu}, we conservatively choose $\mu=3$.
The figure also shows that locally optimizing the curve-fit parameters
helps improve performance. Finally, the figure compares optimization
of Eq.~\eqref{Eq:SystemAx=b} with and without positive-semidefinite
constraints, as well as a simplified model that directly incorporates
$T_1$ and $T_{2\phi}$ parameters instead of a generic block-diagonal
$\beta$. As expected, the learning accuracy improves when the
optimization and model more closely match the ground-truth
Lindbladian.

The overall performance of the learning algorithm depends on the
(number of) initial states, the number of shots per data point, and
the maximum learning depth. Figure~\ref{Fig:Simulations}b shows the
results for a fixed learning depth of 10 for varying shot counts and
number of initial states, sampled uniformly from the set of Pauli
eigenstates. Fixing the number of initial states to 10,
Fig.~\ref{Fig:Simulations}c shows the learning accuracy as a function
of the number of learning depths, given by the maximum learning depth
plus one, used by the protocol.

We briefly consider the setting where measurements remain noiseless,
but the state-preparation error is twice its nominal value, ranging
from 0.3\% to 2.0\%. The results shown in Figs.~\ref{Fig:Simulations}b
and~\ref{Fig:Simulations}c (dashed lines) show that the proposed
learning protocol is robust to state-preparation
errors. In~\sref{SI:StatePrepErrors} we explore noise levels up to
$20\times$ the nominal value. In this regime, the initial state starts
to look like a maximally-mixed state, which leads to observable
expectation values that are shrunk towards zero. It follows from
properties of the binomial distribution underlying the sampling, that
such expectation values have a much larger variance and estimating
them to the same accuracy therefore requires an increased shot count.
Measurement errors affect the learning protocol more directly, and we
therefore try to mitigate them. For the simulations we assume that
measurement errors appear as classification errors following an ideal
measurement. Per qubit, misclassification is represented by a
stochastic 2$\times$2 confusion matrix $M$.  For simplicity we
mitigate the readout errors by applying an estimated $M^{-1}$ to the
measured probabilities, even though this naive approach can result in
negative probabilities~\cite{STE2006ABKa}.

Reverting now to nominal SPAM levels, we compute observable values
based on error-mitigated measurements. The resulting learning
accuracy, as a function of learning depth, is shown in
Fig.~\ref{Fig:Simulations}d. Unlike the noiseless setting, the
learning accuracy levels off at certain bias value due to the
inaccuracy in the estimated confusion matrix. These estimation errors
can be due to shot noise in learning $M$ (the solid lines use $10^4$
shots), but even with infinite shots, the estimate can be inaccurate
due to errors in preparing the $\ket{0}$ and $\ket{1}$ states used
during learning.

For optimal performance under limited shot counts, it is important to
make an appropriate trade-off between the number of initial states,
the learning depth, and the number of shots per data point. We explore
this in Fig.~\ref{Fig:Simulations}e, where we plot the learning
accuracy for different combinations sorted by learning depth, which is
the most dominant parameter. For a fixed learning depth the best
results are generally obtained by maximizing the number of initial
states at the expense of the number of shots per data point, in line
with the $\delta$ rates found for Figs.~\ref{Fig:Simulations}b
and~\ref{Fig:Simulations}c.  The optimal learning depth is around 20
to 40.

The dissipative $T_1$ and $T_{2\phi}$ values in the synthetic model
are inversely proportional to the $\beta$ terms, which have a
magnitude in the order of $10^{-5}$ to $10^{-4}$. Clearly, it requires
a considerable number of shots to resolve these terms accurately.
Indeed, in a setting with 20 initial states, a maximum learning depth
of 30, and no SPAM errors (detailed in~\sref{SI:DissipativeTerms}), we
start to recover these terms using around $10^6$ shots per data point.
In the presence of SPAM errors it is challenging to obtain accurate
coefficient estimates due to the bias induced by measurement-error
mitigation.

The predictive properties of the learned models can be tested by
comparing the expectation values of observables not used during
training to the ideal values. Fig.~\ref{Fig:Simulations}f shows the
expectation values of several weight-three Pauli observables for the
initial state
$\vert$-\hspace{1pt}-+1000\hspace{1pt}-\hspace{1pt}-$\rangle$ using
the ideal model (thick faint lines) as well as the models learned with
(dashed) and without (solid) SPAM errors in the simulated data
points. Even with $10^4$ shots, noiseless learning yields a highly
accurate fit that only improves with increased shot counts. Although
the model learned with SPAM errors gives a relatively close
fit, it does not converge to the ideal curve due to the bias caused by
measurement-error mitigation.

\begin{figure*}[th]
\centering
\includegraphics[width=0.995\textwidth]{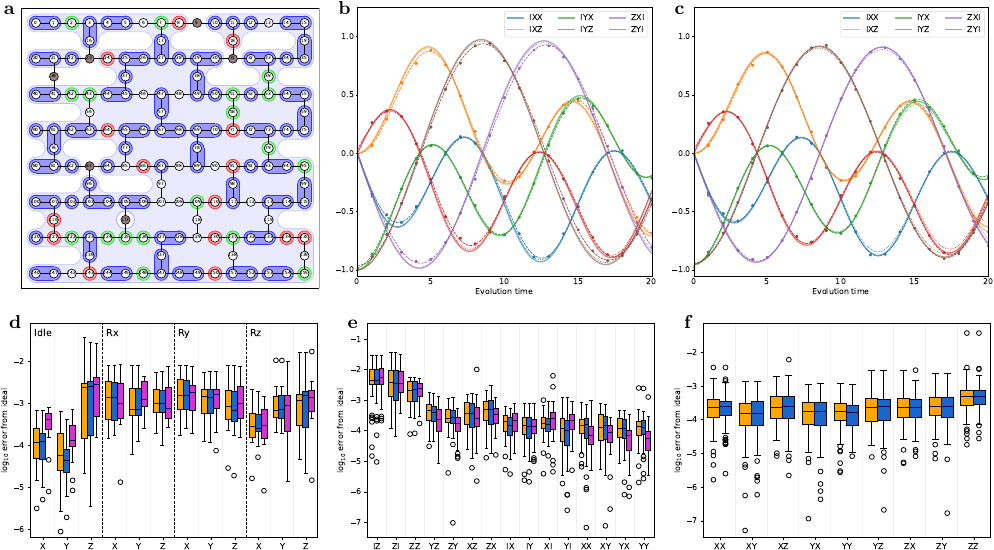}
\caption{(a) Topology of the 156-qubit superconducting quantum
  processor {\it{ibm\_pittsburgh}} used for the experiments along with
  the gates used for Lindblad model learning, including two-qubit Rzz
  and single-qubit Rz gates (blue), Rx gates (red), and Ry gates
  (green). The grey qubits have a SPAM fidelity below 0.97 and are
  excluded from learning; all other qubits (indicated by the light
  blue region) are included in the Lindblad model.  (b) time evolution
  of the learned model (solid) for the initial state $\ket{1r-}$ on
  qubits 93, 94, and 95, along with the error-mitigated data points
  (markers) and the curve fit (dashed). (c) time evolution of the
  fine-tuned model based on unmitigated data (markers). The dotted
  line in plots (b,c) show time evolution of a model learned with
  maximum learning depth 10, respectively without and with fine
  tuning, showing the ability to extrapolate the results in
  time. Plots (d,e) show the distribution of the absolute difference
  between the ideal and the learned Hamiltonian terms for (d) idle
  qubits and single-qubit rotations, and (e) two-qubit Rzz gates for
  the global model (orange), the local model fits (blue) and the
  fine-tuned local models (purple). (f) crosstalk terms on neighboring
  qubit pairs that do not share a gate.}\label{Fig:Experiments}
\end{figure*}

\paragraph{Experiments.}

For the experimental evaluation of the learning protocol we use IBM's
superconducting quantum processor {\it{ibm\_pittsburgh}}, which has
156 qubits arranged in the heavy-hex lattice shown in
Fig.~\ref{Fig:Experiments}a.  As a first step, we analyze the combined
state-preparation and measurement fidelity for each qubit, and select
the qubits with a combined fidelity of at least 0.97. Learning is done
using the TREX readout-error mitigation protocol~\cite{BER2022MTa},
which diagonalizes the readout errors by twirling the measurements
with random X gates (see~\sref{SI:TREX}). We then define a layer of
operations on the selected qubits, consisting of two-qubit ZZ
rotations, single-qubit Pauli rotations, as well as idle time on some
of the qubits. We illustrate the gates in Fig.~\ref{Fig:Experiments}a
and provide all gate details in~\sref{SI:GateLayer}.

We define a Lindblad model on all 150 selected qubits with one- and
two-local Pauli terms for the Hamiltonian and single-qubit Pauli
dissipative terms. The largest connected component of the model has
147 qubits. For learning we generate two sets of nine initial states
such that products of eigenstates of all weight-two Pauli operators
appear on all neighboring qubit pairs~\cite{BER2024Wa}; sampling from
the positive or negative eigenstates is done uniformly at random. For
each initial state we similarly select nine measurement bases such
that each combination of Pauli bases appears once for each pair of
neighboring qubits. Readout errors are mitigated using
TREX~\cite{BER2022MTa}, which requires twirling of the readout. For
this we generate sixteen instances of each circuit with learning
depths up to 20. The circuits for each set of initial states are
executed five times in an interleaved manner between sets. Within each
set, the circuits are executed in a randomized order.  In total we
acquire 7$\cdot 10^4$ shots for readout-error mitigation, 5120 shots
for weight-two Pauli observables and 15,360 shots for weight-one Pauli
observables, since they can be measured in three of the nine
measurement bases.

In experiments, the ground truth is not known, so we cannot directly
evaluate the learning accuracy. However, we can check how well the
time-evolved model fits the data, although this is feasible only for
local models on small subsets of qubits. Fig.~\ref{Fig:Experiments}b
shows time evolution of the model on qubits 93, 94, and 95, along with
the measured data and the curve fits. Despite minimizing
Eq.~\eqref{Eq:SystemAx=b}, there are some noticeable differences
between the evolved model and the data points. In part, this is caused
by not including SPAM errors in the simulated evolution. To improve
the fit, we can augment the (local) model with SPAM terms and fine
tune the model to minimize the misfit to unmitigated data using
forward time evolution to evaluate the misfit, combined with finite
differencing for gradient evaluation (see~\sref{SI:FineTuning} for
more details). Optimization is done with bound constraints on the SPAM
errors. As shown in Fig.~\ref{Fig:Experiments}c, the fine-tuned
results match the experimental data much more closely. The fine tuning
procedure is general and can also be applied to Lindbladians learned
using other protocols.

Finally, we compare the terms in the learned Hamiltonian with those of
the ideal operations. Figures~\ref{Fig:Experiments}d
and~\ref{Fig:Experiments}e show the distribution of absolute errors
for, respectively, single-qubit and two-qubit operations, with and
without fine tuning. Although some of the model inaccuracies are due
to finite shot counts, Pauli-Z errors clearly stand out during for
idle qubits and for two-qubit gates. For crosstalk, shown in
Fig.~\ref{Fig:Experiments}f, the ZZ errors are slightly larger than
the other errors. Plots showing the deviation of the individual model
terms from the ideal Lindbladian can be found in
\sref{SI:LearnedTerms}.

\paragraph{Discussion}

Although the proposed learning method is in principle robust to
state-preparation errors, we have seen that, in practice, these errors
can indirectly affect the learning performance through increased
variance in the observable values and through readout-error
mitigation. The latter is due to the fact that state-preparation
errors affect the learning of the readout errors. Methods that can
characterize readout errors separate from state-preparation errors,
such as~\cite{MAL2026CGSa} are therefore an interesting direction for
future work. The ability to correct readout errors separately would
also obviate the optional model fine-tuning stage, which can currently
be done only on local patches due to its computational complexity.

We remark that separate characterization of state-preparation errors
can also be leveraged by the learning protocol proposed
in~\cite{STI2024MDWa}. Recall that their algorithm forms matrix $A$ in
Eq.~\eqref{Eq:SystemAx=b} based on classically computed observable
values at $t=0$. When estimates for state-preparation errors are
available, these could be included to obtain a better
$A$. Alternatively, the model obtained using~\cite{STI2024MDWa} could
be fine-tuned using our proposed method. This would also mean that
data acquired for the time-evolved state in their protocol is no
longer used merely for curve fitting and gradient evaluation at $t=0$,
but also for fine tuning. State-preparation errors found during fine
tuning could in turn be used to (iteratively) update the system of
equations for better initial models.  Some advantages
of~\cite{STI2024MDWa} are the ease of adding and removing model terms
without necessarily acquiring new data, and that it can be checked
{\it{a priori}} that matrix $A$ has full column rank, since the
entries in $A$ are computed
classically. In~\sref{SI:SimulationImpact2x3} we briefly compare the
performance of both learning methods on a small synthetic problem. A
more thorough comparison between the methods is outside the scope of
the present work.

The proposed fine-tuning algorithm helps mitigate problems with state
preparation but scales exponentially with the number of qubits.  To
better account for crosstalk it is possible to fine tune the local
model in the context of neighboring
qubits~\cite{BAI2019ALa}. Evaluation of the gradient could likely be
improved using backpropagation~\cite{RUM1986HWa}.

A key component of all learning algorithms based on Ehrenfest's
equation is accurate curve fitting. The proposed fitting algorithm
matches the assumed limitation of evolution to successive discrete
time points. In fact, the generalized pencil-of-functions
method~\cite{HUA1989Sa,SAR1995Pa} used in our implementation requires
this. The authors are not aware of algorithms for fitting sums of
exponentially damped sinusoids based on non-uniformly sampled
data. These would clearly be of interest since they would enable more
flexible fitting. In the special case of Hamiltonian learning, there
is no exponential damping and fitting can be done using Fourier
transformations.

An important question in Lindbladian learning is whether the ansatz
includes all relevant terms, and, if not, how to detect this. Given
the measured observable estimates, possibly including high-weight
observable values that were not included in the model learning, the
ideal way to test model accuracy is to perform time evolution and
compare observable expectation obtained from the time-evolved model
with those measured. However, this approach scales exponentially with
the number of qubits and is therefore practical only for small models
or local patches of the model. Another approach, proposed
in~\cite{KRA2025JLOa}, considers the norm $\Vert r\Vert$ of the
residual $r=b - Ax$, as a function of shot count. If the ansatz
contains all relevant terms, it can be expected that the residual norm
continues to decrease as the shot count is increased. Plateaus in the
residual norm can then be seen as an indication that the ansatz is
missing terms (although we have seen that readout errors too can
result in plateaus). Since the learning protocol in~\cite{KRA2025JLOa}
evaluates all entries in the $A$ matrix classically (evaluation only
takes place at $t=0$), it is possible to add model terms purely in
post processing. This is no longer possible when the entries are
measured, since it may require observable expectation values that
cannot be evaluated in any previously used measurement basis.

\paragraph*{Acknowledgments}
The authors thank Alireza Seif, Kristan Temme, and Abhinav Kandala for
helpful discussions.

\paragraph*{Data availability}
Data are available from the authors upon reasonable request.

\onecolumngrid

\clearpage
% --------------------------------------------------------------
% Appendix
% --------------------------------------------------------------
\onecolumngrid
\appendix

\begin{center}
\Large Supplemental Information
\end{center}

\section{Background}\label{SI:Background}

Evolution of a state in an open quantum system can be described by the
Lindblad master equation
$\dot{\rho} = \mathcal{L}(\rho)$~\cite{LIN1976a,GOR1976KSa} with
Lindbladian $\mathcal{L}$. In case the
Lindbladian is time independent we can write it as
\begin{equation}
\mathcal{L}(\rho) = -\frac{i}{\hbar} [H,\rho] + \sum_{ij} \beta_{ij}\left(P_i\rho
P_j^{\dag} - \half\{P_j^{\dag}P_i,\rho\}\right),
\end{equation}
with Pauli operators $P_{\ell}$, Hamiltonian $H = \sum_k \alpha_k P_k$
and positive-semidefinite matrix $\beta$ capturing the coefficients
for the dissipative terms. In this work we develop and study a
protocol for learning the coefficients $\alpha_k$ and $\beta_{i,j}$
from experimental observations.

There are several ways to approach Lindblad learning. The first of
these is based on the matrix logarithm~\cite{ZHA2015Sa,ONO2023KCa}:
the operator of interest $T=\exp(\tau\mathcal{L}(\cdot))$ (we will
assume some unit evolution time $\tau=1$ for simplicity) can be
represented in Liouville space as a transfer matrix
$\bar{\mathcal{T}}$ whose elements can be measured. The transfer
matrix itself can be expressed in the form of a matrix exponential
$\exp{\bar{\mathcal{G}}}$, with generator matrix $\bar{\mathcal{G}}$
from which the Lindbladian terms are readily determined.  Given
$\bar{\mathcal{T}}$ it would seem straightforward to obtain
$\bar{\mathcal{G}}$ by taking the principal branch of the matrix
logarithm of $\bar{\mathcal{T}}$. However, this choice may not be the
correct one~\cite{BOU2003HPCa} and the approach may require searching
over the different branches, as done in~\cite{ONO2023KCa}. Even if the
correct branch could be selected, this approach requires the full
matrix representation of $\bar{\mathcal{T}}$, and therefore scales
exponentially in the number of qubits.

A second approach uses the simulation of state evolution under the
parameterized Lindbladian combined with numerical optimization over
the parameters to minimize an appropriate objective function. Examples
of this approach are the minimization of the negative log-likelihood
of the observed measurements~\cite{SAM2022GBCa,DOB2024CMBa};
minimization of the Kullback-Leibler divergence between the predicted
and experimentally measured measurement
distribution~\cite{BEN2020SAOa}; nonlinear least-squares minimization
of the observable expectation values~\cite{WAN2024La}, and
minimization of the distance between time-evolved Lindbladian and
measured propagators~\cite{BOU2003HPCa,HOW2006TWGa}. Physicality of
the learned Lindbladian in the form of complete positivity can be
enforced using constrained optimization, by post-processing, or
encouraged by means of penalty
functions~\cite{BOU2003HPCa,SAM2022GBCa,WAN2024La}. Although this
approach is quite flexible, it does require time-evolution of the
state under the Lindbladian, which scales exponentially in the number
of qubits, even if the problem is reduced to state-vector
simulation~\cite{WAN2024La}. Moreover, the optimization problems in
this approach are generally non-convex and may therefore require a
good initial solution to avoid convergence to a local minimum.

The third approach, introduced in~\cite{ZUB2021YLBa} in the context of
Hamiltonian learning and subsequently used for Lindblad learning
in~\cite{STI2024MDWa,OLS2024KKKa}, is motivated by Ehrenfest's
theorem. In particular, for an observable $O$ it holds that
\[
\frac{d}{dt}\langle O\rangle
 = \frac{d}{dt}\Tr(O\rho(t)) = \Tr\left(O\frac{d}{dt}\rho(t)\right)
=\langle O\mathcal{L}(\rho(t))\rangle.
\]
Expanding with $H = \sum_j \alpha_j P_j$ and simplifying gives
\begin{equation}\label{Eq:ddtO}
\frac{d}{dt}\langle O\rangle
=
\sum_j \alpha_j \langle -i[O,P_j]\rangle
 + \sum_{ij}\sfrac{\beta_{ij}}{2}(
\langle P_j^{\dag}[O,P_i]\rangle + \langle [P_j^{\dag},O]P_i\rangle),
\end{equation}
where the expectation values are with respect to
$\rho(t) = e^{t\mathcal{L}}\rho(0)$.  For a known initial state
$\rho(0)$ we can evaluate the expectation values appearing in the
right-hand side of Eq.~\eqref{Eq:ddtO} and obtain a weighted sum of
parameters $\alpha$ and $\beta$. Repeating this for a set of randomly
selected initial states $\rho_k$ and observables $O_k$ gives a linear
system of equations of the form $Ax=b$ where $x$ contains the Lindblad
parameters $\alpha$ and $\beta$ and $b$ represents a vector of
observable derivatives $\frac{d}{dt}\Tr(O_k\rho_k(0))$. The parameter
values can then be found by solving a least-squares problem,
minimizing $\half\Vert Ax-b\Vert_2^2$ over $x$.  Evaluation of the
observable gradient can be done using forward
finite-differencing~\cite{ZUB2021YLBa}:
\[
\frac{d}{dt}\langle O\rangle_{\rho(0)} = \frac{\langle
  O\rangle_{\rho(\delta t)} - \langle O\rangle_{\rho(0)}}{\delta t} +
\mathcal{O}(\delta t),
\]
which only requires the experimental estimation of
$\langle O\rangle_{\rho(\delta t)}$, since
$\langle O\rangle_{\rho(0)}$ can be evaluated classically for a known
$\rho(0)$. To obtain a high-accuracy estimate of the gradient, this
approach may require a small $\delta t$.  As an alternative,
\cite{STI2024MDWa} proposes evaluation of
$y_k = \langle O\rangle_{\rho(t_k)}$ at a series of time points $t_k$,
followed by low-order polynomial fitting of the $(t_k, y_k)$ data
points using robust polynomial interpolation~\cite{KAN2017KPa}. The
polynomial fit can then be differentiated to obtain an estimate of the
gradient at $t=0$.

\section{Proposed learning protocol}\label{SI:ProposedProtocol}

One of the disadvantages of the approach proposed
in~\cite{STI2024MDWa}, is that the observable values at various
evolution times $t$ are used only to fit the curve, whose gradient is
then evaluated only at $t=0$. This seems wasteful, and in this work we
therefore propose to use the gradient information at all measured
times. In terms of the system of linear equations, this means that
vector $b$ now contains one entry per data point of the measured
observable. Each row in matrix $A$ now corresponds to the observable
evaluated at evolution time $t$. Unlike~\cite{STI2024MDWa}, which
classically computes the observable values at $t=0$, we measure the
entries in $A$ for all required $t\geq 0$. For a fixed Pauli
observable $O$ we can measure $O(t) = \Tr(O\rho(t))$ and perform curve
fitting and differentiation to obtain the left-hand side of
Eq.~\eqref{Eq:ddtO}. For the right-hand side of the equation, consider
a single Hamiltonian term:
\[
\langle [O,P_j]\rangle = \Tr([O,P_j]\rho(t)) = \Tr((OP_j - P_jO)\rho(t))
\]
When $O$ and $P_j$ commute we have $[O,P_j]=0$, which means that the
coefficient for $\alpha_j$ in the current equation is zero. Otherwise,
it holds that $P_jO = -OP_j$ and $P_jO = \pm iQ$ for some Pauli
operator $Q$. In this case, we can therefore obtain the coefficient
for $\alpha_j$ by measuring $\langle Q\rangle = \Tr(Q\rho(t))$.  A
similar derivation applies for the dissipative terms, resulting in the
coefficients for $\beta_{i,j}$. Each observable $O$ thus results in a
set of related observables that must be measured. By appropriately
selecting observables $O$, many of these related observables can be
reused in other entries in the system of equations.

For the evaluation of the gradients we also apply curve fitting. As
part of our other contributions in this paper, we show in
section~\ref{SI:DampedSinusoids} that the observable values
$\langle O(t)\rangle$ can be expressed as sums of exponentially damped
sinusoids. Section~\ref{SI:PolynomialFits} then shows that these
curves cannot in general be fit using low-order polynomials. In
Section~\ref{SI:CurveFitting}, we therefore propose an algorithm for
fitting the measured data using sums of exponentially damped
sinusoids.

At the time of writing we became aware of~\cite{AGU2025WDRa}, which
proposes a similar protocol based on time-evolved states. It differs
from ours in that is uses finite-differencing to evaluate
gradients. Note that, for each gradient evaluation, this requires
measuring the observable value at two or more closely spaced time
points.

\section{Observable values as a function of time}%
\label{SI:ObservableValues}\label{SI:DampedSinusoids}

The Lindblad master equation describes the evolution of a density
matrix in open quantum systems. Given an initial density matrix
$\rho(0)$ and assuming a time-independent Lindbladian, we can write
the density matrix at time $t$ as
$\rho(t) = \exp(t\mathcal{L})\rho(0)$. Measuring the expectation value
of an observable $O$ at the different evolution times gives a
real-valued function $\langle O(t)\rangle = \Tr(\rho(t) O)$. For our
learning procedure, we need to fit a curve through values of
$\langle O(t)\rangle$ sampled at a given set of time points.  In this
section, we therefore study the form of $\langle O(t)\rangle$.  We
first express the superoperator corresponding to $\mathcal{L}$ as
$\mathscr{L} = \sum_{k}\lambda_k \kett{\zeta_k}\bbra{\xi_k}$, with
left and right eigenvectors $\bbra{\xi_k}$, $\kett{\zeta_k}$ and
corresponding eigenvalue
$\lambda_k$. Following~\cite[Eq.~(205)]{GYA2020a}, we then express the
density matrix evolution in Liouville space as
\[
\kett{\rho(t)} = \sum_{k=1}^{d^2}e^{t\lambda_k}\kett{\zeta_k}\bbrakett{\xi_k}{\rho(0)}.
\]
Vectorizing the observable to
$\bbra{O}$ we obtain
\[
\langle O(t)\rangle = \bbrakett{O}{\rho(t)} = 
\sum_{k=1}^{d^2}e^{t\lambda_k}\bbrakett{O}{\zeta_k}\bbrakett{\xi_k}{\rho(0)}.
\]
The term $c_k = \bbrakett{O}{\zeta_k}\bbrakett{\xi_k}{\rho(0)}$ is a
time-independent constant, and we therefore have
\[
\langle O(t)\rangle = \sum_k c_k e^{t\lambda_k},
\]
where $\exp(t\lambda_k)$ represents a (complex) damped sinusoid. Since
$\langle O(t)\rangle$ is a real function it follows that we can write
$\langle O(t)\rangle$ as the sum of real-valued damped sinusoids:
\begin{equation}\label{Eq:ExpOt}
\langle O(t)\rangle = \sum_j a_j e^{b_j t}\cos(\omega_j t + \varphi_j),
\end{equation}
with coefficients $a_j, b_j, \omega_j, \varphi_j \in \mathbb{R}$.

In practice we can expect the observable to be affected by
state-preparation and measurement errors. This results in an initial
state $\rho'(0)$ that differs from the ideal $\rho(0)$. However, since
no assumptions were made on $\rho(0)$, the functional form of
$\langle O(t)\rangle$ in Eq.~\eqref{Eq:ExpOt} is unaffected by any
state-preparation errors. As discussed in more detail in
Section~\ref{SI:TREX}, we perform twirled measurements, which
diagonalize the Pauli-Z errors and therefore result in a fixed scaling
factor for each Pauli-Z observable $O$. This scaling factor can be
included in the $a_j$ terms. Any basis changes prior to measurement in
the computational basis can be absorbed in $\kett{\zeta_k}$, and
readout error therefore does not affect the form of function.

\section{Curve fitting}\label{SI:CurveFitting}\label{SI:CurveFittingProcedure}

For our Lindblad learning procedure we are given a collection of
expectation values for different observables and evolved states at
successive integer multiples of some unit evolution time. Fitting the
data points of each available observable and initial state by a sum of
damped sinusoids of the form
\begin{equation}\label{Eq:Curve}
f(t) = \sum_j a_j e^{b_j t}\cos(\omega_j t + \varphi_j),
\end{equation}
can be done using the classic Prony's
method~\cite{PRO1795a,PER2010Sa}. However, this method is sensitive to
noise, and a more stable approach is given by the generalized
pencil-of-functions (GPOF) method~\cite{HUA1989Sa,SAR1995Pa}. One
disadvantage of using GPOF on noisy data, is that it generally fits
curves through the data points and therefore results in curves that
overfit the measured data. To avoid this, we modify the algorithm and
postprocess the initial fit as follows.

The input to the GPOF algorithm is a set of $y$ values at regularly
sampled intervals. These are converted into matrices and successively
transformed by means of a singular-value decomposition, an
eigendecomposition, and a linear least-squares optimization problem
(see~\cite{SAR1995Pa} for details). For the singular-value
decomposition we limit the range of singular values, truncating terms
below $10^{-9}$ of the largest singular value. The eigendecomposition
results in individual eigenvalues as well as in conjugate pairs. We
identify these conjugate pairs up to some numerical precision to
facilitate further processing. After solving the least-squares
problem, the curve fit to the data is of the form
$f(t) = \sum_j r_j z_j^t$, where $r_j$ and $z_j$ are complex scalars.
We have found it beneficial to compute the misfit of the curve to the
data $\sum_k (y_k - f(k))^2$ and to repeat the GPOF on the residual
vector $v_k = y_k - f(k)$, if needed. The resulting $r$ and $z$ terms
are then added to the original terms to obtain an augmented set of
terms.

The next stage of the fitting procedure discards terms that damp too
quickly, that is, we remove terms for which $\vert z_j\vert$ is too
small. Such terms are essentially zero after a few unit time steps and
generally appear only to fit low-depth outliers, resulting in large
and incorrect gradient approximations at those depths. Optionally, we
filter out high-frequency components (i.e., where the argument
$\mbox{arg}(z_j)$ is too large).  Combining conjugate pairs, where
needed, we arrive at a curve representation of the form
Eq.~\eqref{Eq:Curve}

At this point, the curve generally still overfits the data. We
therefore attempt to select a minimal subset of terms that fits the
data to within a preselected multiplicative factor $\mu$ of the
expected misfit of the data for the given shot counts (we will derive
these quantities in Sec.~\ref{SI:ExpectedMisfit}). For the selection
of the terms we use a greedy algorithm that iteratively adds one or
two terms to the selected set, optionally followed by local
optimization of the parameters.  In particular, at each iteration, the
greedy algorithm checks which single term would reduce the misfit to
the data the most when added to the selected set. In some cases it was
found that a principal component of the curve, for instance
$0.3e^{-t}$, was formed by adding two terms, say $0.7e^{-t}$ and
$-0.4e^{-t}$. In such cases, both terms, would marginally decrease or
even increase the misfit when considered individually, but greatly
reduce the misfit when taken together.  The greedy algorithm therefore
considers the addition of pairs of terms with appropriate conditions
on the required reduction of the misfit, compared to that of a single
term, to avoid unnecessary pairwise addition.

Finally, after each update of the selected set of terms, we locally
solve a non-convex least-squares problem, minimizing the $\ell_2$
distance between the curve and the measured data points, to fine tune
the coefficients of the selected terms (if the optimization process
fails to improve the fit to the data we use the original
coefficients). Optimization can be done using gradient-descent based
methods~\cite{NOC2006Wa}, warm-started from the original coefficients
of the selected terms. Whenever the resulting misfit falls below the
misfit threshold, or when no more terms are available for addition we
return the current fit as the solution. Otherwise, we reset the
coefficients of the selected terms to their original value and proceed
with the next iteration of the greedy selection algorithm to add more
terms.

As an aside, we remark that curve fitting in~\cite{STI2024MDWa} is
done using low-degree polynomials. This may suffice for the evaluation
of the gradient at time $t=0$ when data is available only for small
depths. For larger depths, however, the approach clearly cannot be
expected to work, since the data may exhibit too many oscillations
(see also Sec.~\ref {SI:PolynomialFits}). In addition, at least for
theoretical analysis, the authors of~\cite{STI2024MDWa} assume that
(continuous) evolution times are sampled from a Chebyshev or uniform
measure. Such a scheme may be difficult to implement in practice,
since it would require observable estimation at arbitrary evolution
times. By contrast, we assume access only to the operation obtained by
unit time evolution. As a consequence, by repeating the operation
multiple times if needed, we only have access to observable values at
discrete times given as integer multiples of the unit evolution time.

\begin{figure}[t]
\centering
\begin{tabular}{cc}
\begin{minipage}{0.35\textwidth}
\begin{center}
\includegraphics[width=\textwidth]{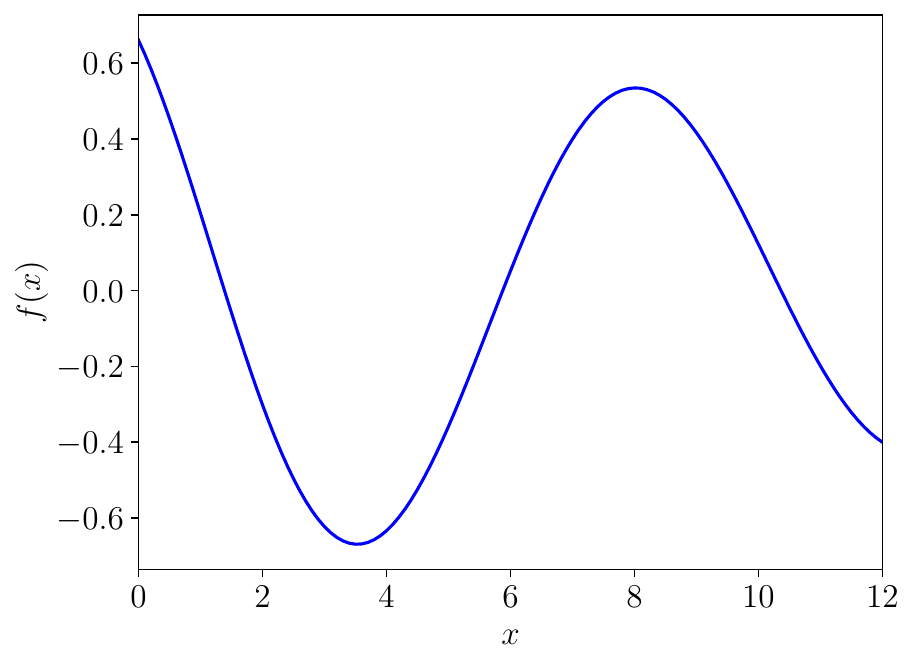}%
\begin{picture}(0,0)
\put(-21,100){({\bf{a}})}
\end{picture}
\\
\includegraphics[width=\textwidth]{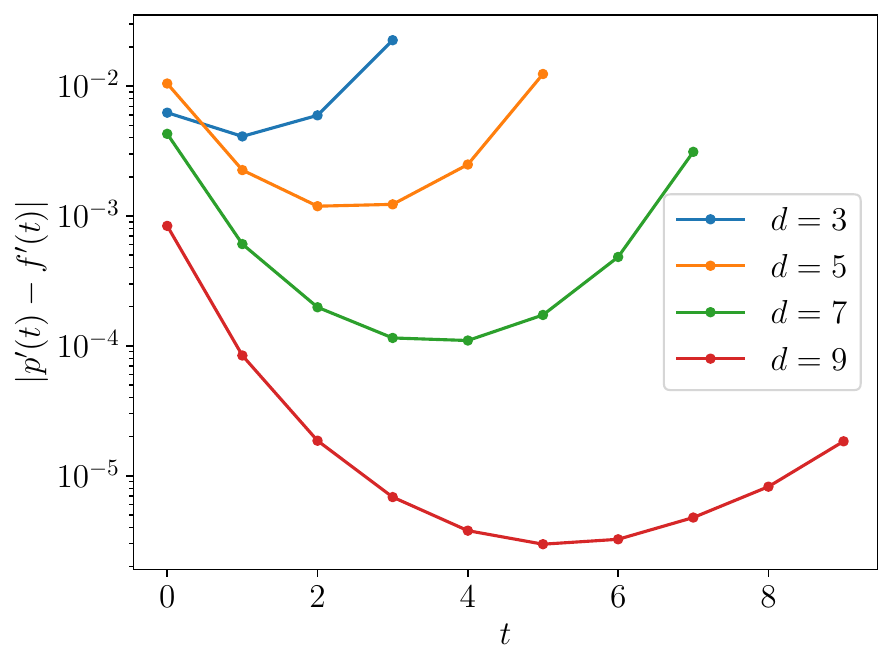}%
\begin{picture}(0,0)
\put(-20,102){({\bf{b}})}
\end{picture}
\end{center}
\end{minipage}
&
\begin{minipage}{0.58\textwidth}
\begin{center}
\includegraphics[width=\textwidth]{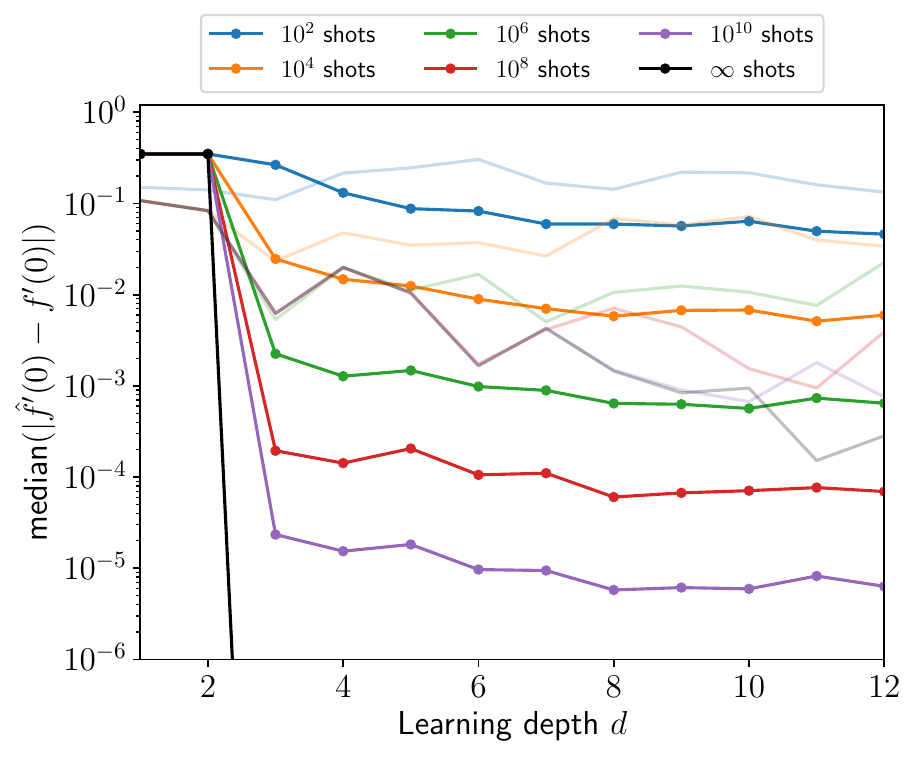}\\[-2pt]
({\bf{c}})
\end{center}
\end{minipage}
\end{tabular}
\caption{Plot of (a) $f(t) = 0.8\exp(-0.05t)\cos(0.7t + 0.6)$, (b) the
  absolute difference between $f'(t)$ and the derivative $p'(t)$ of
  the $d$-th order polynomial fit $p(t)$ obtained from exact data
  points $f(t)$ evaluated at $t=[0,1,\ldots,d]$. The dots represent
  the absolute difference at the data point times; we connect the dots
  by lines for visual reference. The gradient misfit is largest at the
  end points. (c) The median absolute error in the gradient estimate
  based on fitting a sum of exponentially damped sinusoids (solid with
  dots) and the best polynomial fit (faint lines) for different
  learning depths and shot counts when sampling the data, based on 100
  trials per setting.}\label{Fig:PolynomialFit}
\end{figure}

\section{Comparison with polynomial fits}\label{SI:PolynomialFits}

We now highlight the need to fit the observable values
$\langle O(t)\rangle$ using a sum of exponentially damped sinusoids
rather than polynomials. Consider a simple example where the
observable expectation value evolves as
$f(t) = 0.8\exp(-0.05t)\cos(0.7t + 0.6)$, as illustrated in
Figure~\ref{Fig:PolynomialFit}(a). Suppose we can measure $f(t)$
exactly at $t = [0,1,\ldots,d]$ for some maximum learning depth
$d$. We can then fit a $d$-th order polynomial through the data points
to obtain the fit $p(t)$. In Figure~\ref{Fig:PolynomialFit}(b) we plot
the error in the gradient estimation $\vert p'(t) -
f'(t)\vert$. Increasing the maximum learning depth leads to improved
estimates, but the gradient misfit at the end points (including $t=0$)
remains quite large, even under the assumption of exact evaluation of
$f(t)$.

Since $f(t)$ can be fit exactly using a single exponentially damped
sinusoids, we now consider the performance of fitting using this class
of functions. To make the setting more interesting we now limit the
number of shots that can be used to estimate the observable
expectation values $f(t)$. Fitting using the procedure described
earlier, we obtain the results shown in
Figure~\ref{Fig:PolynomialFit}(c). Starting at a learning depth $d=3$
we see that the absolute error in the gradient estimate scales
according to the standard limit $\mathcal{O}(1/\sqrt{n})$ in the
number of shots $n$ and slightly decreases with increased maximum
learning depth. For comparison we also determine the lowest-degree
polynomial that fits the data up to the exact misfit between the ideal
and sampled data (this quantity would generally not be known in
practice). The median gradient misfit at $t=0$ for different learning
depths (over 100 instances of shot samples) is shown by the faint
lines in Figure~\ref{Fig:PolynomialFit}(c). In this case, gradient
estimation using polynomial fits is clearly much less accurate than
the fit using a sum of damped sinusoids. For completeness, we note
that, if $f(t)$ were a polynomial, the polynomial fit would naturally
excel.

\section{Expected deviation from the ideal curve}\label{SI:ExpectedMisfit}

The expectation value $\langle O\rangle$ for a given Pauli observable
$O$ is usually experimentally estimated by averaging individual
samples or shots with value $\pm 1$.  We can model sampling of an
ideal observable value $v$ by means of a scaled binomial distribution
$B(n,p)$, where $n$ is the number of samples and $p=(1+v)/2$ is the
probability of measuring $1$. Letting $Y\sim B(n,p)$, we define the
random variable associated with the observable value by
$V = 2Y/n - 1$. The mean and variance of $V$ follow from
$\mathbb{E}(Y) = np$ and $\mathrm{Var}(Y) = np(1-p)$ and are given by
$\mathbb{E}(V) = 2p-1 = v$ and
$\mathrm{Var}(V) = 4p(1-p)/n = (1-v^2)/n$, respectively. Given the
empirically measured observable value $\hat{v}$ we can estimate the
variance as $\hat{\sigma}^2 = (1-\hat{v}^2)/n$.  We want to avoid
variance estimates close to zero when $\vert \hat{v}\vert$ is close to
one and $n$ is small. To achieve this, we impose a small minimum
variance proportional to $1/n$.

With sufficiently many samples, the observable random variable $V$
that describes the deviation of $\hat{v}$ from its mean $v$, is well
approximated with a normal distribution $\mathcal{N}(v,
\sigma^2)$. Since we only consider the deviation from $v$, we shift
$V$ to have zero mean and write
$V_0 = V - \mathbb{E}(V) \sim \mathcal{N}(0,\sigma^2)$. The
distribution of $Y_1 = \vert V_0\vert$, which closely models
$\vert\hat{v} - v\vert$, follows a half-normal distribution with
\[
\mathbb{E}(Y_1) = \sigma\sqrt{2/\pi},\qquad
\mathrm{Var}(Y_1) = \sigma^2\left(1 - \frac{2}{\pi}\right).
\]
The squared difference $Y_2 = V_0^2$ approximates $(\hat{v} - v)^2$
and is described by a stretched chi-squared distribution $\chi_1^2$, with
mean and variance
\begin{align*}
\mathbb{E}(Y_2) & = \mathbb{E}(V_0^2) = \mathrm{Var}(V_0) - \mathbb{E}(V_0)^2\\
\mathrm{Var}(Y_2) &= \mathbb{E}(Y_2^2) - \mathbb{E}(Y_2)^2 = \mathbb{E}(V_0^4)
- \mathbb{E}(V_0^2)^2.
\end{align*}
The $k$-th central moment for the normal distribution is given by
$\sigma^k(k-1)!!$, where $k!!$ denotes the double factorial. For
$k=4$ we have $(k-1)!! = 3$, and therefore
\[
\mathbb{E}(Y_2) = \sigma^2,\qquad
\mathrm{Var}(Y_2) = 3\sigma^4 - \sigma^4 = 2\sigma^4.
\]
By substituting $\sigma$ with the empirical $\hat{\sigma}$ we can
obtain a reasonable estimate of the expected absolute or squared
difference between each data point and ideal curve. Summing over the
data points we can then obtain the overall expected difference and its
variance. The expected difference can be used during curve fitting as
a reference to prevent overfitting or underfitting the data.

\section{Nyquist-rate sampling}\label{SI:Nyquist}

In Section~\ref{SI:DampedSinusoids}, we saw that observable values, as
a function of evolution time, can be written as a sum of damped
sinusoids. For curve fitting, we are given a series of observable
expectation values at integer multiples of a given unit evolution
time. Disregarding any shot noise, note that, in order to reconstruct
the function from such equispaced samples, we need to ensure that the
sampling rate exceeds the so-called Nyquist rate. This rate is equal
to twice the largest frequency in the given function and therefore
imposes limits on the magnitude of the coefficients in the
Lindbladian, which is reflected in the frequencies appearing in the
observable value, or on the length of the unit evolution time, which
corresponds to the sampling rate. Reducing the unit evolution time of
the Lindbladian describing a certain operation effectively amounts to
implementing a fractional operation.

\subsection{Addition of frequencies}\label{SI:FrequencyAddition}

As a first example, consider the Hamiltonian for a three-qubit system
corresponding to an $\alpha$-fractional CX operation on the first two
qubits, conditioned on the second qubit, and a $\beta$-fractional X
gate on the third qubit:
\[
H = \alpha\sfrac{\pi}{4}(III - XII - IZI + XZI) + \beta\sfrac{\pi}{2}(IIX - III),
\]
such that the overall operation amounts to $U(t) = \exp(-itH)$. For a
given initial state $\rho$ and observable $O$ we have
\[
\langle O(t)\rangle := \Tr(U(t)\rho U^{\dag}(t)O).
\]
Now, consider the action of $U(t)$ on the initial state
$\rho := \ket{r,-,l}\bra{r,-,l} = \sfrac{1}{8}(I+Y)\otimes
(I-X)\otimes (I-Y)$, where $r$ and $l$ are respectively the positive
and negative eigenstates of $Y$. For observable $O = IXY$ it can be
verified that
\[
\langle{IXY(t)}\rangle = \sfrac{1}{2}(\cos(\alpha\pi
  t)+1)\cdot\cos(\beta\pi t),
\]
where the first and second term respectively come from the action of
the fractional CX and X gates. Rewriting this using the trigonometric
identity
$\cos(\theta)\cos(\varphi) =
(\cos(\theta-\varphi)+\cos(\theta+\varphi))/2$ gives
\[
\langle{IXY(t)}\rangle = \sfrac{1}{4}\left(\cos((\alpha-\beta)\pi
 t)+ \cos((\alpha+\beta)\pi t)\right) + \sfrac{1}{2}\cos(\beta\pi t).
\]
Assuming nonnegative $\alpha$ and $\beta$, it can be seen that the
highest frequency of the signal is $(\alpha+\beta)/2$. As mentioned
above, in order to reconstruct the function from equispaced samples,
we need to sample at a rate exceeding the Nyquist rate, which is twice
this frequency.  For a unit evolution time of one, we acquire samples
at integer time values, which therefore requires $\alpha+\beta < 1$
for successful fitting. For larger values of $\alpha+\beta$, we cannot
hope to recover the function $\langle IXY(t)\rangle$ based on the
sampled data.

\subsection{Generalization}

More generally, consider an $n$-qubit example where the Hamiltonian is
given by $H = \sfrac{\theta}{2} \sum_{i=1}^n X_i$, such that the unit
time evolution amounts to applying an $R_x(\theta)$ gate on each
qubit.  For a zero initial state and an all-Z observable
$O = Z^{\otimes n}$, it is easy to see that
\[
\langle Z^{\otimes n}(t)\rangle = \prod_{i=1}^{n}\cos(\theta t) =
\cos^n(\theta t) = \frac{1}{2^m}\sum_{k=0}^{m} {m\choose
  k}\cos\big((1-m+2k)\theta t\big),
\]
with $m=n-1$. The largest frequency component in the observable
function is equal to $n\theta/2\pi$, although it should be noted that
the weights of the frequencies away from the center $k\approx m/2$
decrease rapidly, since most of the weight of
$\sfrac{1}{2^m}{m\choose k}$ is centered around that point.

The appearance of high-frequency components, relative to the sampling
rate, means that there are practical limitation on Hamiltonian and
Lindblad learning methods that rely on curve fitting. For instance, it
imposes certain maximum rotation angles for the gates under
consideration. In our proposed algorithm we limit the computation of
$d\langle O(t)\rangle/dt$ to weight-one observables to keep the number
of learning bases to a minimum, and therefore never fit curves for
higher-weight observables. In general, higher-weight observables could
provide additional information and may be used in other algorithms,
such as the one proposed by Stilck Fran\c{c}a {\it et
  al.}~\cite{STI2024MDWa}. In this latter case, the limitations on the
largest frequency component of the signal still apply, although the
effects of sampling below Nyquist rate may be somewhat damped due to
the fact that the gradient is evaluated only at $t=0$.

\subsection{Weight-one observable}

We now show that sampling rate restrictions are not limited to
higher-weight observables. Indeed, they can appear in somewhat
surprising ways even in weight-one Pauli observables. For instance,
consider the circuit from Sec.~\ref{SI:FrequencyAddition} with initial
state $\ket{r,-,l}$, and weight-one observable $O = IXI$. We now have
\[
\langle IXI(t)\rangle = -\frac{1  + \cos(\alpha \pi t)}{2} =
-\cos^2\left(\frac{\alpha\pi t}{2}\right).
\]
The Nyquist rate here is given by $\alpha$, and sampling at integer
multiples of the evolution time of a full CX gate therefore coincides
exactly with the Nyquist rate. This situation changes rather
dramatically when adding a ZZ interaction between the second and third
qubits as $H_{\delta} \equiv \gamma (\pi/2) IZZ$ (recall we have an X
gate on qubit three). The exact expression for $\langle IXI(t)\rangle$
is now given by
\[
\langle IXI(t)\rangle =\frac{\cos \left(\frac{\alpha \pi t}{2}\right) \left\{-\cos \left(\frac{\alpha \pi t}{2}\right) \left[\beta ^2+\gamma ^2 \cos \left(\sqrt{\beta ^2+\gamma ^2}\pi t\right)\right]-\beta  \gamma  \sin \left(\frac{\alpha \pi t}{2}\right) \left[\cos \left(  \sqrt{\beta^2+\gamma ^2} \pi t \right)-1\right]\right\}}{\beta ^2+\gamma ^2}.
\]
In the weak-$\gamma$ regime, this can be simplified up to the lowest
order as
\[
\langle IXI(t)\rangle \approx -\cos^2 \left(\frac{\alpha \pi
    t}{2}\right) + \frac{\gamma}{2\beta} \sin(\alpha \pi
t)\left[1-\cos\left(\sqrt{\beta^2+\gamma^2}\pi t\right)\right].
\]
Therefore, with the added ZZ interaction, the expectation value
possesses an additional weak-amplitude oscillation, but with a larger
Nyquist rate of $\alpha + \sqrt{\beta^2+\gamma^2} >
\alpha+\beta$. When sampling at a rate above $\alpha$, but below
$\alpha+\beta$, we therefore cannot accurately recover this component
of the observable value function, which results in a loss of accuracy
in the gradient of the observable value.

\section{Simulations}

\subsection{Synthetic test problems}\label{SI:SyntheticSetup}

\begin{figure}[th]
\centering\setlength{\tabcolsep}{10pt}
\begin{tabular}{ccc}
\raisebox{4pt}{\includegraphics[width=0.27\textwidth]{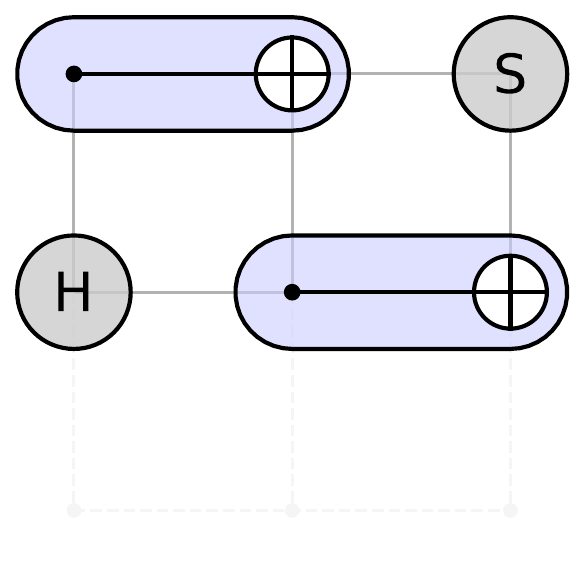}}&
\raisebox{4pt}{\includegraphics[width=0.27\textwidth]{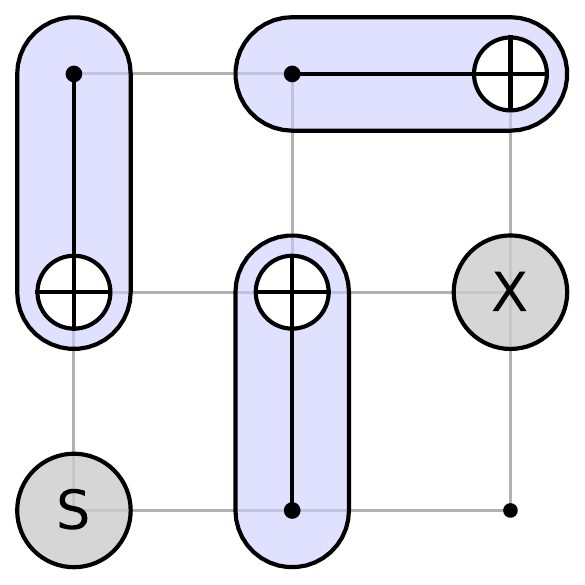}}&
\includegraphics[width=0.315\textwidth]{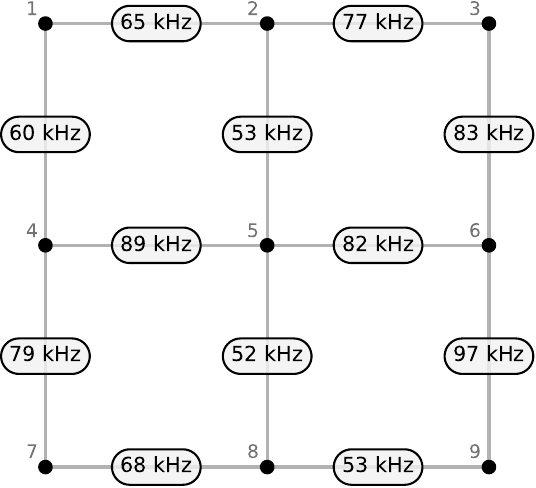}\\
({\bf{a}}) & ({\bf{b}}) & ({\bf{c}})\\[10pt]
\multicolumn{3}{c}{\includegraphics[width=0.9\textwidth]{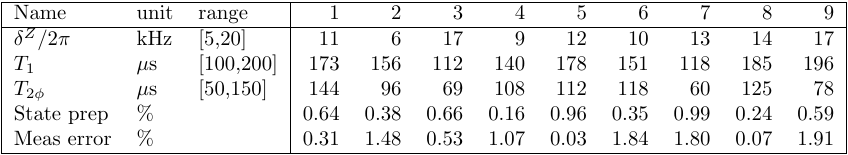}}\\[2pt]
\multicolumn{3}{c}{({\bf{d}})}
\end{tabular}
\caption{(a) 6-qubit and (b) 9-qubit circuits on a hypothetical
  3$\times$3 quantum processor. The Hamiltonian corresponding to the
  gates is scaled such that unit-time evolution is equal to 30\% of
  the overall gate duration for the 6-qubit case and 20\% for the
  9-qubit case. We assume a uniform gate time of 50ns. (c) ZZ
  interactions between the qubits, and (d) a summary of the
  single-qubit properties including Pauli-Z errors, $T_1$ and
  $T_{2\phi}$ times, and state-preparation and readout
  errors.}\label{SI:FigTestProblems}
\end{figure}

For our simulations we use a hypothetical quantum processor inspired
by superconducting devices. The qubits on the devices follow a regular
$3\times 3$ grid topology and interactions are restricted to
neighboring qubits. The ground-truth Lindbladian is given by
\begin{equation}\label{Eq:ModelL}
 \mathcal{L}[\rho]  = - i [H_g +
 H_{\delta},\rho] + \mathcal{D}[\rho],
\end{equation}
where $H_g$ denotes the Hamiltonian for the ideal operation and
$H_{\delta}$ describes the deviation from the ideal Hamiltonian due to
single-qubit $Z$ and two-qubit $ZZ$ errors on neighboring qubit pairs
(S):
\[
H_{\delta} = \sum\limits_{j} \frac{\delta_{j}^{Z}}{2} Z_{j} + \sum\limits_{(j,k)\in\mathcal{S}} \frac{\delta_{jk}^{ZZ}}{2} Z_{j}Z_{k}.
\]
Finally, we write the dissipator term in Eq.~\eqref{Eq:ModelL} as
$\mathcal{D} = \mathcal{D}_{\downarrow} + \mathcal{D}_{\phi}$, where
\begin{align*}
\mathcal{D}_{\downarrow}[\rho] &= \sum_{j}\beta_{\downarrow j}\Big(
\sigma_j^{-}\rho\sigma_j^{+} - \half\{\sigma_j^{+}\sigma_j^{-},\rho\}\Big)\\
\mathcal{D}_{\phi}[\rho] &= \sum_{j}\sfrac{\beta_{\phi}}{2}\left(
Z_j\rho Z_j^{\dag} - \rho\right)
\end{align*}
respectively represent amplitude damping and pure dephasing, with
$\sigma_{j}^{\pm} = \half(X_j \mp iY_j)$ (see
also~\cite{BRE2002Pa,SCH2024SLa,MAL2025SPGa}).  The $T_1$ and
$T_{2\phi}$ lifetimes are related to the decay rates as
$\beta_{\downarrow} = 1/T_1$ and $\beta_{\phi}=1/T_{2\phi}$.  We can
alternatively express the dissipator in terms of single-qubit Pauli
terms and a block-diagonal $\beta$ matrix, where each block
corresponds to a single qubit, and where the block $\beta_j$ for qubit
$j$ is given by
\begin{equation}\label{Eq:BetaBlock}
\beta_j =
\begin{blockarray}{cccc}
& X_j & Y_j & Z_j \\
\begin{block}{c(ccc)}
X_j  & \frac{\beta_{\downarrow j}}{4} & -i\frac{\beta_{\downarrow j}}{4} & 0 \\
Y_j  & i\frac{\beta_{\downarrow j}}{4} & \frac{\beta_{\downarrow j}}{4} & 0 \\
Z_j  & 0 & 0 & \frac{\beta_{\phi j}}{2}\\
\end{block}
\end{blockarray}\ .
\end{equation}

The Hamiltonian terms $H_g$ are chosen such that evolving by the gate
time implements a desired layer of gates. We consider two layers of
gates, shown in Fig.~\ref{SI:FigTestProblems}. We assume square pulses
and a uniform gate time fixed at $\tau_g = 50$ns. For instance, for a
$\mbox{CX}_{\theta}$ gate on qubits 1 and 2, we have
$H_g = \sfrac{\omega}{2}(Z_1 + X_2 - Z_1X_2)$, where $\omega$ is
chosen such that $\omega\tau_g = \theta$. For a standard CX gate with
$\theta=\pi/2$, this amounts to setting $\omega$ to $2\pi\times 5$MHz.
For specific problem instances we select $\delta_{j}^{Z}$ and
$\delta_{jk}^{ZZ}$ uniformly at random from the interval
$2\pi\times [5,20]$ kHz and $2\pi\times [50,100]$ kHz,
respectively. We further choose the qubit $T_1$ and $T_{2\phi}$
coherence times uniformly at random from the intervals $[100,200]$
$\mu$s and $[50,150]$ $\mu$s, as summarized in
Fig.~\ref{SI:FigTestProblems}. For the $2\times 3$ circuit we
disregard terms outside of the selected region and therefore
effectively operate on a 6-qubit topology.

For the evaluation of the learning protocol, we choose a unit
evolution time of $\tau$ that corresponds to some fraction of the gate
time $\tau_g$, namely 30\% for the $2\times 3$ setting, and 20\% for
the $3\times 3$ setting. By doing so, we can avoid introducing
systematic curve fitting errors due to sampling below the Nyquist
rate, as described in more detail in Section~\ref{SI:Nyquist}, and
therefore work within the fundamental limitations of the protocol.

\subsection{Selection of the curve fitting tolerance}\label{SI:ChoiceMu}

\begin{figure}[!b]
\centering
\begin{tabular}{ccc}
% Script: simulations/fig_noiseless_2x3c.py
\includegraphics[width=0.32\textwidth]{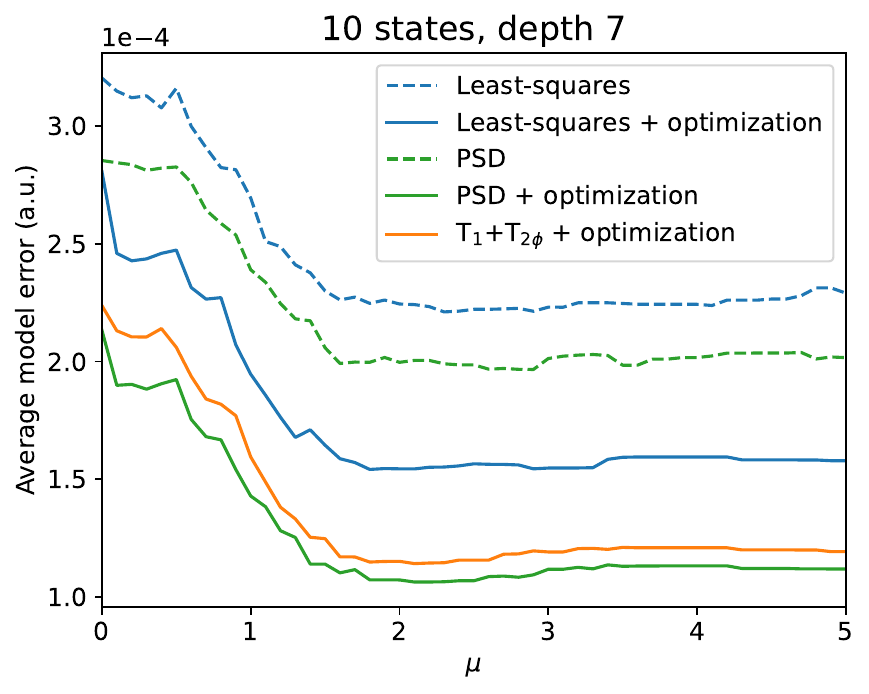}&
\includegraphics[width=0.32\textwidth]{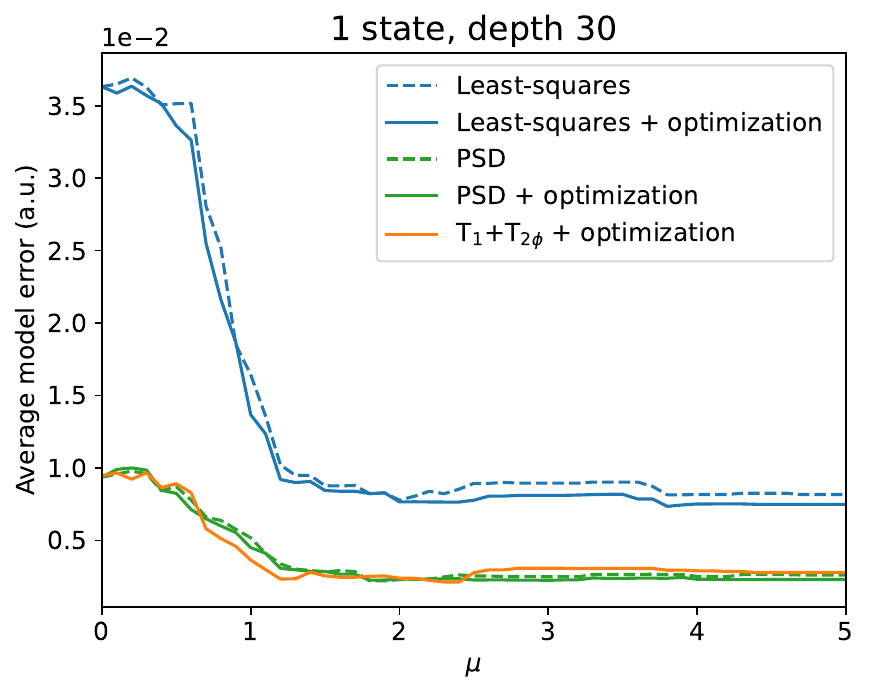}&
\includegraphics[width=0.32\textwidth]{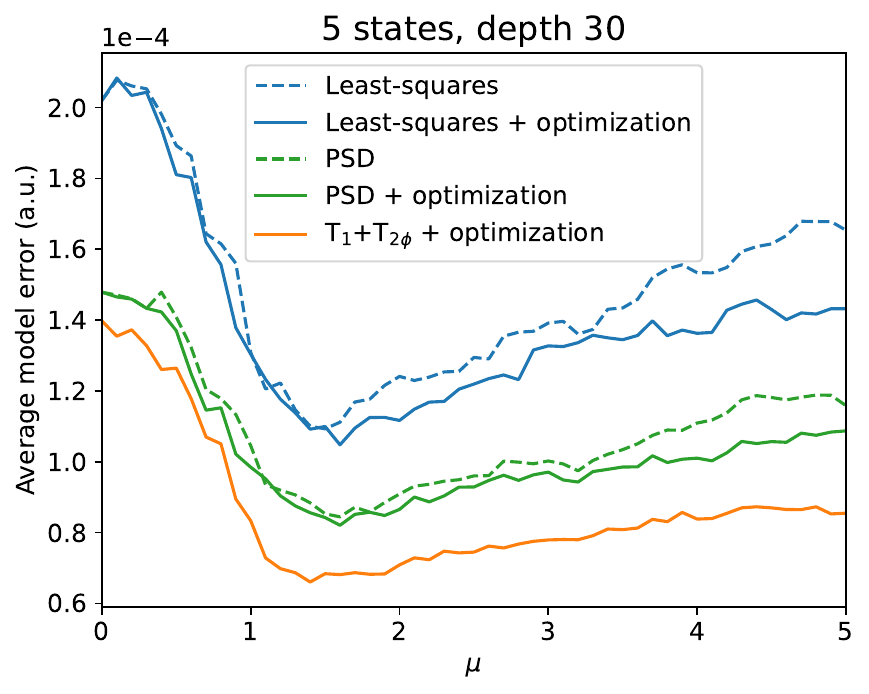}\\
% Script: simulations/fig_noiseless_3x3b.py
\includegraphics[width=0.32\textwidth]{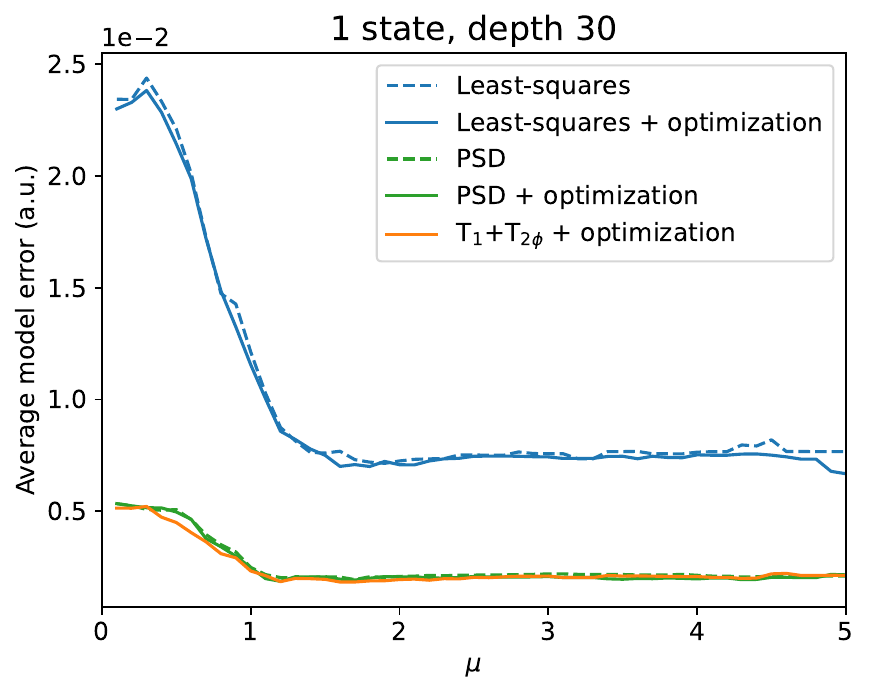}&
\includegraphics[width=0.32\textwidth]{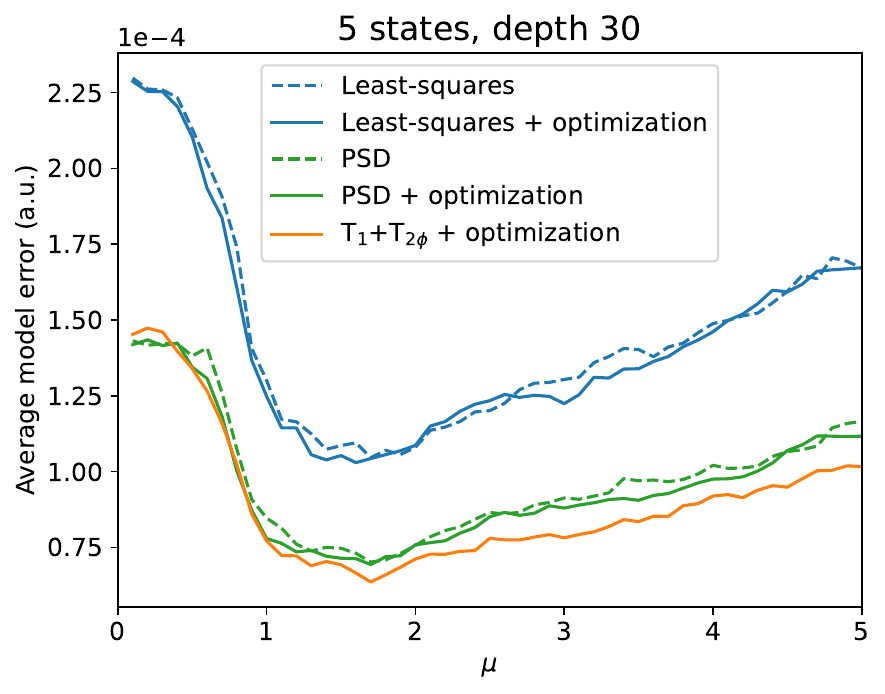}&
\includegraphics[width=0.32\textwidth]{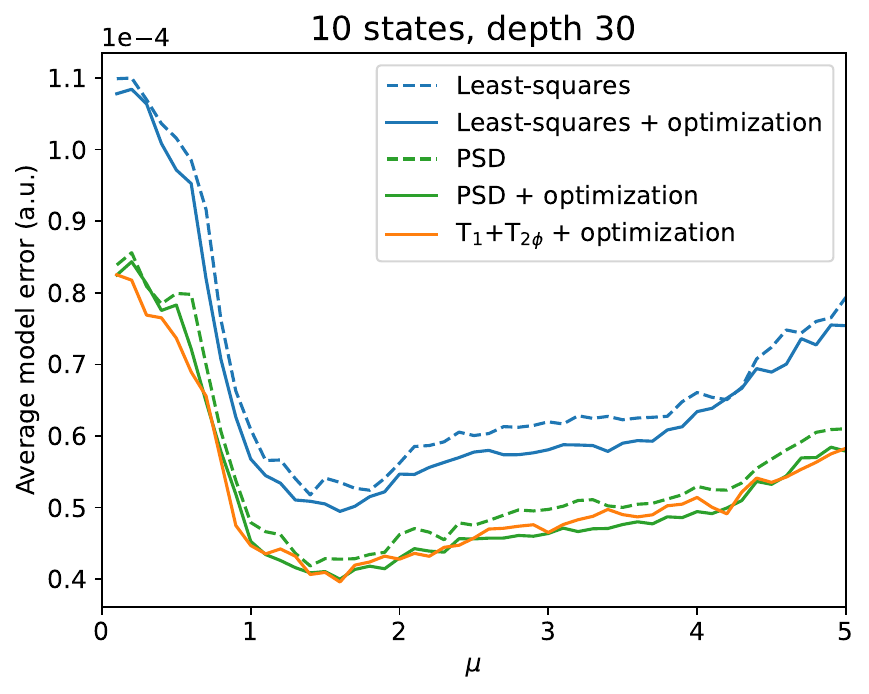}
\end{tabular}
\caption{Dependence of the learned model parameters on the
  curve-fitting tolerance parameter $\mu$ and different
  optimization methods for (top row) the 2$\times$3 circuit and
  (bottom row) the 3$\times$3 circuit. The title above each plot
  indicates the number of initial states and the maximum learning
  depth. The model accuracy is computed as the geometric mean of the
  median model term errors obtained with $10^k$ shots for integer
  values $k \in [2,8]$.}\label{SI:FigMultiplier}
\end{figure}

Fitting of the curves is done with a target residual that is defined
as a multiple $\mu$ of the expected deviation derived in
Section~\ref{SI:ExpectedMisfit}. Here, we study the impact the choice
of $\mu$ has on the accuracy of the learned model. We consider both
the 6-qubit and 9-qubit test problems and use different methods for
fitting the model, namely, minimizing the system of equations residual
$\half\Vert Ax -b\Vert_2^2$ without (least-squares) or with
positive-semidefinite constraints (PSD) based on curve fits with or
without additional local optimization. We also consider a specialized
model that directly parameterizes the $T_1$ and $T_{2\phi}$ terms,
rather than use the more generic $\beta$ matrix. In
Fig.~\ref{SI:FigMultiplier} we plot the resulting model accuracy as a
function of multiplier $\mu$. The model accuracy is computed as the
geometric mean of the median model parameter error obtained for data
with $10^k$ shots for integer values $k \in [2,8]$. This allows us to
choose a parameter that simultaneously works well for different shot
counts.  Local optimization of the curve fits shows improved results
compared to directly using the curve terms obtained from the
generalized pencil-of-functions (GPOF) method. Moreover, we see that
the results obtained using PSD constraints are better than those
obtained using unconstrained least-squares minimization. Using the
specialized $T_1$-$T_{2\phi}$ model gives the best results; perhaps
unsurprisingly, since this setting best matches the synthetic
Lindbladian.

In all setting, the best choice of $\mu$ lies between $1.5$ and
$2$. Performance generally degrades rapidly for $\mu < 1.5$ and more
gradually for $\mu > 2$. For the remainder of the paper we select
$\mu=3$ as a conservative choice. Note that these results are obtained
based on simulation data without state-preparation and measurement
errors.

\subsection{Impact of curve fitting and limited shot counts}\label{SI:SimulationImpact2x3}

Even in the absence of state-preparation and measurement errors the
simulated data will deviate from the ideal observable values due to
finite shot counts and the numerical time evolution of the density
matrix by means of the Trotter-Suzuki product formula. In this section
we explore how these errors affect the learned Lindblad model.

We focus on the 6-qubit test problem, which is still small enough to
enable an explicit matrix representation of the superoperator
$\mathscr{L}$ of the Lindbladian. This allows us to evaluate the time
evolution of the state based on the eigendecomposition of
$\mathscr{L}$, which also allows us to determine the (numerically)
exact expectation value curves $\langle O(t)\rangle$ for selected
observables. In particular, we can compute the eigendecomposition
$\mathscr{L} = B\Lambda B^{-1}$, where
$\Lambda = \mbox{diag}(\lambda)$ contains the eigenvalues and the
columns $b_i$ in $B$ represent the corresponding left eigenvectors. We
then have
\[
\exp(\tau \mathscr{L}) = B\ \mbox{diag}(\exp(\tau \lambda))B^{-1},
\]
where exponentiation of the eigenvalues is applied elementwise. For an
initial state $\rho_0$ and observable $O$ we then have
\[
\langle O(\tau)\rangle = \bbra{O}\exp(\tau\mathscr{L})\kett{\rho_0}
= \bbra{O}B\,\mbox{diag}(\exp(\tau\lambda)) B^{-1}\kett{\rho_0}.
\]
Expansion of the matrix-vector terms and noting that
$\exp(\tau\lambda_i)$ is nothing but the sum of a real and an
imaginary weighted damped sinusoid, it is easy to see that we can
write $\langle O(\tau)\rangle$ as a sum of damped sinusoids with known
coefficients. Since we know that the expectation value is real, we can
discard all the terms whose weight is not real-valued (see also
Sec.~\ref{SI:DampedSinusoids} for more details).

\begin{table}[!th]
\centering
% Source: simulations/fig_noiseless_2x3c.py
\begin{tabular}{rp{5.9cm}llll}
\hline
{\small{\#}} & Setting & $H$ & $H^c$ & $D$ & $D^c$ \\
\hline
{\small{1}} & Exact data, exact curve             & $5.00\cdot 10^{-15}$ & -- & $4.47\cdot 10^{-15}$ & -- \\
{\small{2}} & Exact data, fitted curve            & $3.05\cdot 10^{-8}$ & $1.04\cdot 10^{-7}$ & $2.73\cdot 10^{-8}$ & $3.49\cdot 10^{-8}$ \\
\hline
{\small{3}} & Simulated data ($\infty$), exact curve & $5.91\cdot 10^{-11}$ & -- & $8.60\cdot 10^{-13}$ & -- \\
{\small{4}} & Simulated data ($\infty$), fitted curve & $3.04\cdot 10^{-8}$ & $1.04\cdot 10^{-7}$ & $2.63\cdot 10^{-8}$ & $3.47\cdot 10^{-8}$ \\
{\small{5}} & Exact data, fitted curve ($\infty$) & $3.04\cdot 10^{-8}$ & $1.04\cdot 10^{-7}$ & $2.63\cdot 10^{-8}$ & $3.47\cdot 10^{-8}$ \\
\hline
{\small{6}} & Simulated data ($10^2$), exact curve & $4.22\cdot 10^{-2}$ & $6.76\cdot 10^{-2}$ & $1.40\cdot 10^{-2}$ & $6.13\cdot 10^{-3}$ \\
{\small{7}} & Simulated data ($10^2$), fitted curve & $4.89\cdot 10^{-2}$ & $7.90\cdot 10^{-2}$ & $1.49\cdot 10^{-2}$ & $9.54\cdot 10^{-3}$ \\
{\small{8}} & Exact data, fitted curve ($10^2$)   & $1.33\cdot 10^{-2}$ & $2.90\cdot 10^{-2}$ & $7.61\cdot 10^{-3}$ & $3.16\cdot 10^{-3}$ \\
\hline
{\small{9}} & Simulated data ($10^4$), exact curve & $1.77\cdot 10^{-3}$ & $5.01\cdot 10^{-3}$ & $1.30\cdot 10^{-3}$ & $1.31\cdot 10^{-3}$ \\
{\small{10}} & Simulated data ($10^4$), fitted curve & $7.55\cdot 10^{-3}$ & $9.13\cdot 10^{-3}$ & $3.76\cdot 10^{-3}$ & $1.25\cdot 10^{-3}$ \\
{\small{11}} & Exact data, fitted curve ($10^4$)   & $6.46\cdot 10^{-3}$ & $6.41\cdot 10^{-3}$ & $3.07\cdot 10^{-3}$ & $5.81\cdot 10^{-4}$ \\
\hline
{\small{12}} & Simulated data ($10^6$), exact curve & $1.37\cdot 10^{-4}$ & $6.95\cdot 10^{-4}$ & $1.51\cdot 10^{-4}$ & $1.89\cdot 10^{-4}$ \\
{\small{13}} & Simulated data ($10^6$), fitted curve & $6.03\cdot 10^{-4}$ & $9.69\cdot 10^{-4}$ & $3.94\cdot 10^{-4}$ & $2.52\cdot 10^{-4}$ \\
{\small{14}} & Exact data, fitted curve ($10^6$)   & $5.32\cdot 10^{-4}$ & $7.24\cdot 10^{-4}$ & $3.89\cdot 10^{-4}$ & $2.31\cdot 10^{-4}$ \\
\hline
{\small{15}} & Simulated data ($10^8$), exact curve & $1.55\cdot 10^{-5}$ & $6.00\cdot 10^{-5}$ & $3.40\cdot 10^{-5}$ & $1.62\cdot 10^{-5}$ \\
{\small{16}} & Simulated data ($10^8$), fitted curve & $3.49\cdot 10^{-5}$ & $8.31\cdot 10^{-5}$ & $6.76\cdot 10^{-5}$ & $3.41\cdot 10^{-5}$ \\
{\small{17}} & Exact data, fitted curve ($10^8$)   & $2.68\cdot 10^{-5}$ & $6.87\cdot 10^{-5}$ & $4.42\cdot 10^{-5}$ & $3.08\cdot 10^{-5}$ \\
\hline
\end{tabular}
\\
\caption{Difference between the exact and the learned Lindbladians for
  different settings of the 6-qubits test problem without
  state-preparation and measurement errors. All settings use 20
  initial states and a maximum learning depth of 30. The settings
  either use exact data points based on the eigendecomposition of the
  superoperator, or data generated by means of Trotter-Suzuki time
  evolution. The curves are either the exact curves based on the
  eigendecomposition, or fitted curves based on the sampled data. The
  number in brackets denotes the number of shots per data
  point.}\label{Table:ModelFitIdeal}
\vspace*{20pt}
\centering
% Source: simulations/fig_noiseless_2x3c.py
\begin{tabular}{rp{5.9cm}llll}
\hline
{\small{\#}} & Setting & $H$ & $H^c$ & $D$ & $D^c$ \\
\hline
{\small{1}} & Exact data, exact curve             & $6.08\cdot 10^{-15}$ & -- & $8.61\cdot 10^{-15}$ & -- \\
{\small{2}} & Exact data, fitted curve            & $2.14\cdot 10^{-6}$ & $5.57\cdot 10^{-6}$ & $1.16\cdot 10^{-5}$ & $3.31\cdot 10^{-6}$ \\
%\hline
{\small{3}} & Simulated data ($\infty$), fitted curve & $2.25\cdot 10^{-6}$ & $5.54\cdot 10^{-6}$ & $1.17\cdot 10^{-5}$ & $3.57\cdot 10^{-6}$ \\
\hline
{\small{4}} & Simulated data ($10^2$), fitted curve & $2.05\cdot 10^{-1}$ & $2.10\cdot 10^{-1}$ & $5.62\cdot 10^{-2}$ & $4.86\cdot 10^{-2}$ \\
%\hline
{\small{5}} & Simulated data ($10^4$), fitted curve & $1.71\cdot 10^{-2}$ & $2.75\cdot 10^{-2}$ & $6.09\cdot 10^{-3}$ & $4.16\cdot 10^{-3}$ \\
%\hline
{\small{6}} & Simulated data ($10^6$), fitted curve & $5.23\cdot 10^{-3}$ & $8.45\cdot 10^{-3}$ & $1.77\cdot 10^{-3}$ & $1.60\cdot 10^{-3}$ \\
%\hline
{\small{7}} & Simulated data ($10^8$), fitted curve & $5.09\cdot 10^{-4}$ & $1.47\cdot 10^{-3}$ & $6.42\cdot 10^{-4}$ & $3.94\cdot 10^{-4}$ \\
\hline
\end{tabular}
\\
\caption{Results obtained by processing the data used for
  Table~\ref{Table:ModelFitIdeal} using the learning protocol by
  Stilck Fran\c{c}a {\it{et al.}}~\cite{STI2024MDWa}, with the
  original polynomial curve fitting replaced by our fitting
  method. These simulations have 20 initial states and a maximum
  learning depth of 30.}\label{Table:ModelFitIdealSF}
\end{table}

Table~\ref{Table:ModelFitIdeal} shows the one-norm of the errors in
the learned model terms grouped by the error on the ideal (non-zero)
Hamiltonian terms $H$; the one-norm of learned Hamiltonian terms not
appearing in the ideal model, $H^c$, which should ideally be zero; and
likewise for the dissipative terms $D$ and their complement $D^c$. For
these simulations we have no state-preparation and measurement
errors. When using exact observable data and exact observable curves,
the results are exact up to numerical precision. Replacing the exact
curves by the fitted curves increases the error to roughly
$10^{-8}$. Simulating the data using a sixth-order Trotter-Suzuki
scheme with 100 steps per unit evolution time combined with exact
curves shows a simulation error of order $10^{-11}$. From the next two
rows we see that, in the infinite shot limit, the error due to curve
fitting limits the final model accuracy to around $10^{-8}$. Next, we
impose shot counts on the simulated observable data. With $10^2$ shots
per data point the model accuracy is around $10^{-2}$. Increasing the
number of shots to $10^8$ improves this to $10^{-5}$. Scaling of the
error in terms of the number of shots $n$ roughly follows the
$\mathcal{O}(1/\sqrt{n})$ standard quantum limit.

We also applied a slightly modified version of the learning protocol
by Stilck Fran\c{c}a {\it{et al.}}~\cite{STI2024MDWa} (SF) on the same
data set: time-series data for all one- and two-local observables
$O$. For each $O$ we fit a curve through the corresponding time-series
data (using our fitting method rather than the original low-degree
polynomial fits used in~\cite{STI2024MDWa}) and evaluate
$\sfrac{d}{dt}\!\left.\langle O\rangle\right\vert_{t=0}$. In the
system of equations $Ax=b$, this gives a single entry in vector
$b$. The entries in the corresponding row in $A$ are classically
computed based on the given initial state (these entries are exact up
to numerical precision since we did not include state-preparation or
readout errors in this simulation). We then fit the model with a
positive semi-definite constraint on the block-diagonal matrix
$\beta$. The results, shown in Table~\ref{Table:ModelFitIdealSF}, are
slightly worse than those obtained using our method
(Table~\ref{Table:ModelFitIdeal}). This comparison does have a number
of caveats. First of all, the maximum circuit depth used for curve
fitting may not be optimal for the SF approach. Using a smaller depth
and more shots per data point may improve the results. Second, given
the importance of the gradient at $t=0$, it might make sense to
allocate more shots to data points for smaller depths. Overall, more
work is needed to evaluate the relative performance of the two
methods.

\begin{table}[!th]
\centering
% Script: simulations/fig_noiseless_2x3c.py
\begin{tabular}{lrrrrrrcrrrrrr}
\hline
Number of shots & \multicolumn{6}{c}{$T_1$} && \multicolumn{6}{c}{$T_{2\phi}$}\\
\hline
$10^{2}$ shots                      & $\infty$ &   4 &  30 & $\infty$ &  32 & $\infty$ & & $\infty$ & $\infty$ &  22 &   6 &   8 &   5 \\
$10^{3}$ shots                      & 164 & $\infty$ & 280 &  49 &  69 & $\infty$ & &  55 & $\infty$ &  27 & 1153 &  81 &  14 \\
$10^{4}$ shots                      & 170 & 123 & 248 &  68 & 220 &  62 & & $\infty$ &  59 &  47 & 104 &  33 &  59 \\
$10^{5}$ shots                      & 637 &  85 & 236 & 231 & 333 &  76 & &  54 &  85 &  42 &  68 &  63 &  99 \\
$10^{6}$ shots                      & 173 & 152 & 121 & 122 & 184 & 215 & & 1348 & 100 &  74 &  85 &  79 & 113 \\
$10^{7}$ shots                      & 187 & 160 & 117 & 153 & 185 & 142 & & 138 &  95 &  74 & 110 & 108 & 137 \\
$10^{8}$ shots                      & 175 & 158 & 113 & 138 & 179 & 155 & & 148 &  97 &  69 & 111 & 115 & 114 \\
$\infty$ shots                      & 173 & 156 & 112 & 140 & 178 & 151 & & 144 &  96 &  69 & 108 & 112 & 118 \\
Ideal value                         & 173 & 156 & 112 & 140 & 178 & 151 & & 144 &  96 &  69 & 108 & 112 & 118 \\
\hline
\end{tabular}
\\[1pt]
({\bf{a}})\\[8pt]
\begin{tabular}{lrrrrrr}
\hline
Number of shots & \multicolumn{6}{c}{$\beta_{\downarrow}$}\\
\hline
$10^{2}$ shots                      & 0.00e+00 & 8.66e-04 & 1.25e-04 & 0.00e+00 & 1.16e-04 & 0.00e+00 \\
$10^{3}$ shots                      & 2.29e-05 & 0.00e+00 & 1.34e-05 & 7.70e-05 & 5.44e-05 & 0.00e+00 \\
$10^{4}$ shots                      & 2.20e-05 & 3.05e-05 & 1.51e-05 & 5.52e-05 & 1.71e-05 & 6.05e-05 \\
$10^{5}$ shots                      & 5.88e-06 & 4.44e-05 & 1.59e-05 & 1.62e-05 & 1.12e-05 & 4.92e-05 \\
$10^{6}$ shots                      & 2.16e-05 & 2.47e-05 & 3.10e-05 & 3.08e-05 & 2.04e-05 & 1.74e-05 \\
$10^{7}$ shots                      & 2.01e-05 & 2.35e-05 & 3.21e-05 & 2.45e-05 & 2.03e-05 & 2.64e-05 \\
$10^{8}$ shots                      & 2.14e-05 & 2.38e-05 & 3.30e-05 & 2.72e-05 & 2.10e-05 & 2.42e-05 \\
$\infty$ shots                      & 2.17e-05 & 2.40e-05 & 3.35e-05 & 2.68e-05 & 2.11e-05 & 2.48e-05 \\
Ideal value                         & 2.17e-05 & 2.40e-05 & 3.35e-05 & 2.68e-05 & 2.11e-05 & 2.48e-05 \\
\hline
\end{tabular}
\\[1pt]
({\bf{b}})\\[8pt]
\begin{tabular}{lrrrrrr}
\hline
Number of shots & \multicolumn{6}{c}{$\beta_{\phi}$}\\
\hline
$10^{2}$ shots                      & 0.00e+00 & 0.00e+00 & 3.47e-04 & 1.29e-03 & 9.76e-04 & 1.43e-03 \\
$10^{3}$ shots                      & 1.36e-04 & 0.00e+00 & 2.81e-04 & 6.50e-06 & 9.23e-05 & 5.21e-04 \\
$10^{4}$ shots                      & 0.00e+00 & 1.26e-04 & 1.60e-04 & 7.21e-05 & 2.25e-04 & 1.27e-04 \\
$10^{5}$ shots                      & 1.38e-04 & 8.87e-05 & 1.80e-04 & 1.10e-04 & 1.19e-04 & 7.57e-05 \\
$10^{6}$ shots                      & 5.57e-06 & 7.52e-05 & 1.02e-04 & 8.86e-05 & 9.55e-05 & 6.64e-05 \\
$10^{7}$ shots                      & 5.43e-05 & 7.86e-05 & 1.02e-04 & 6.82e-05 & 6.93e-05 & 5.49e-05 \\
$10^{8}$ shots                      & 5.07e-05 & 7.72e-05 & 1.09e-04 & 6.74e-05 & 6.53e-05 & 6.56e-05 \\
$\infty$ shots                      & 5.21e-05 & 7.81e-05 & 1.09e-04 & 6.94e-05 & 6.70e-05 & 6.36e-05 \\
Ideal value                         & 5.21e-05 & 7.81e-05 & 1.09e-04 & 6.94e-05 & 6.70e-05 & 6.36e-05 \\
\hline
\end{tabular}
\\[1pt]
({\bf{c}})
\caption{Recovery of (a) the $T_1$ and $T_{2\phi}$ coherence times,
  and (b,c) the corresponding $\beta$ parameters of the 6-qubit model
  for different shot counts per data point. Here we apply the learning
  protocol with a specialized model that directly parameterizes the
  coherence times and solves the model using least-squares
  minimization with positive-semidefinite constraints on the $\beta$
  matrix. Each column corresponds to one of the six
  qubits.}\label{SI:TableT1T2phi}
\end{table}

\subsection{Recovery of the dissipative terms}\label{SI:DissipativeTerms}

The ground-truth Lindblad model includes dissipative terms that
capture the $T_1$ and $T_{2\phi}$ coherence times. In this section we
study how accurate these values are recovered from simulation data
with finite shot counts for the $2\times 3$ circuit with ideal
state preparation and measurement. The ideal and recovered coherence
times, rounded to the nearest integer are summarized in
Table~\ref{SI:TableT1T2phi}a. A substantial number of shots is needed
to accurately resolve the coherence times. Recall that these times are
inversely proportional to the $\beta$ parameters, and large coherence
times result in small $\beta$ values. Indeed, as seen in
Tables~\ref{SI:TableT1T2phi}b and~\ref{SI:TableT1T2phi}c, the $\beta$
values are quite small. It takes around $10^6$ shots per data point to
see close fits between the ideal and recovered $\beta$ parameters and,
consequently, the coherence times.

\subsection{Simulation of state-preparation and measurement errors}\label{SI:StatePrepErrors}

So far, our simulations have been free from state-preparation and
measurement errors. To better reflect experiments on actual quantum
processors we need to introduce these sources of errors. We model
noisy state preparation of $\rho_0 = \ket{0}\bra{0}$ as
$(1-s)\ket{0}\bra{0} + s\ket{1}\bra{1}$, where $s$ represents the
state preparation error. After applying ideal basis changes, we evolve
the state according to the Lindblad master equation to the target
time. At this point we change to the desired measurement basis using
ideal gates to obtain a final state $\rho$ and simulate noisy
measurement in the computational basis. For this, we start with the
ideal measurement probabilities $p$ of the computational basis states
given by the diagonal elements of $\rho$. A commonly used model for
measurement errors~\cite{GEL2020a,MAC2019ZOa,HAP2012HPa} is the
application of single-qubit assignment errors to the ideal
probabilities:
\[
\tilde{p}_j = M_jp_j = \left[\begin{array}{cc}1-m_j & m_j\\
 m_j& 1-m_j\end{array}\right]p_j,
\]
where each qubit $j$ can have its own measurement error matrix
$M_j$. We assume that the probability of measuring 0 as 1 and
measuring 1 as 0 are both equal to $m$. If needed, this can always be
achieved by means of twirling the readout~\cite{BER2022MTa}, which
symmetrizes the $M_j$ matrices. We obtain the measurement
probabilities over all qubits as $\tilde{p} = (\bigotimes_j M_j)p$.

In order to mitigate the errors, we need to learn the $M_j$
matrices. Ideally, this would be done by determining the probability
of measuring 0 or 1 when starting from the $\ket{0}$ state. In
practical settings, we only have access to noisy initial states and a
limited shot count, which means that we can only estimate
$\hat{M}_j \approx M_j$. For mitigation, we
follow~\cite{STE2006ABKa,KAN2017MTTa} and obtain our estimate
$\hat{p}$ of probability vector $p$ by applying
$\bigotimes_j \hat{M}_j^{-1}$ on the measured $\tilde{p}$. Note,
however, that this direct inversion may result in unphysical results
in the sense that entries in $\hat{p}$ may be
negative~\cite{NAC2020UJBa}.

In the implementation of the classical simulation we consider both
finite and infinite shot counts. When taking infinite shots we
directly compute $\hat{p} = (\bigotimes_j \hat{M}_j^{-1}M_j)p$. When
the shot count is finite, we first need to simulate the acquisition of
individual shots, which we put in a vector $q$ of length $2^n$,
containing the empirical probability distribution. We then compute
$(\bigotimes_j \hat{M}_j^{-1})q$ to obtain $\hat{p}$. Given $\hat{p}$,
we can apply a Hadamard transformation to determine the estimated
Pauli-Z expectation values.

In Table~\ref{Table:AmplifiedSPE} we consider the effect that
amplified state-prepration has on learnability of the Lindblad terms
for the synthetic $2\times 3$ problem described in
Table~\ref{SI:FigTestProblems}. We multiply the canonical
state-prepration error by factors between 0 (no state-preparation
error) and 20 (state prepration errors ranging from 3.2\% to 19.2\%),
but assume there are no readout errors. The results in
Table~\ref{Table:AmplifiedSPE} show that while the learning protocol
is robust to state-preparation errors, increased state-preparation
errors by themselves do affect the overall model accuracy. In the
present simulation setting, increasing the state-preparation error
results in an initial state that increasingly resembles the
maximally-mixed state. This causes the observable expectation values
to shrink towards zero (for the maximally-mixed state all non-identity
observables would have a zero expectation value). Sampling of
observable expectation values that are close to zero have a larger
variance than values close to $\pm 1$, requiring a larger shot count
to achieve the same overall variance. Indeed, at the infinite sampling
limit, there is no noticeable difference between the different levels
of state-preparation error.

\begin{table}
% simulations/revision_2x3c.py
\begin{tabular}{p{3.9cm}lllll}
\hline
Setting & $0\times$ & $0.02\times$ & $0.2\times$ & $2\times$ & $20\times$ \\
\hline
Simulated data ($10^2$) & $5.22\cdot 10^{-2}$ & $4.38\cdot 10^{-2}$ & $4.39\cdot 10^{-2}$ & $5.18\cdot 10^{-2}$ & $9.14\cdot 10^{-2}$ \\
Simulated data ($10^4$) & $7.77\cdot 10^{-3}$ & $8.24\cdot 10^{-3}$ & $7.66\cdot 10^{-3}$ & $9.49\cdot 10^{-3}$ & $9.01\cdot 10^{-3}$ \\
Simulated data ($10^6$) & $6.22\cdot 10^{-4}$ & $6.07\cdot 10^{-4}$ & $5.54\cdot 10^{-4}$ & $7.14\cdot 10^{-4}$ & $1.18\cdot 10^{-3}$ \\
Simulated data ($10^8$) & $3.74\cdot 10^{-5}$ & $3.76\cdot 10^{-5}$ & $3.58\cdot 10^{-5}$ & $3.50\cdot 10^{-5}$ & $5.18\cdot 10^{-5}$ \\
Simulated data ($\infty$) & $2.99\cdot 10^{-8}$ & $3.04\cdot 10^{-8}$ & $2.70\cdot 10^{-8}$ & $3.23\cdot 10^{-8}$ & $4.79\cdot 10^{-9}$ \\
Exact data            & $2.99\cdot 10^{-8}$ & $3.10\cdot 10^{-8}$ & $3.61\cdot 10^{-8}$ & $3.22\cdot 10^{-8}$ & $4.04\cdot 10^{-9}$ \\
Exact data, exact curve             & $8.21\cdot 10^{-15}$ & $7.56\cdot 10^{-15}$ & $9.00\cdot 10^{-15}$ & $1.18\cdot 10^{-14}$ & $8.26\cdot 10^{-15}$ \\
\hline
\end{tabular}
\caption{The Hamiltonian error for the synthetic $2\times 3$ problem
  for different amplification factors of the canonical
  state-preparation error (see Table~\ref{SI:FigTestProblems}) and no
  readout errors. Except for the last row, gradients of the observable
  expectation values are determined based on curve fitting. The last
  row uses the exact curve based on the eigendecomposition of the
  superoperator representation of the
  Lindbladian.}\label{Table:AmplifiedSPE}
\end{table}

\section{Experiments}

\subsection{Layer definition and circuit preparation}\label{SI:GateLayer}

For the experimental validation of the learning protocol we define a
layer of operations on a 156-qubit chip. We select qubits that have an
overall SPAM fidelity exceeding 97\%.  We add two-qubit Rzz gates and
single-qubit rotation gates on some of the selected qubits, and leave
the remaining selected qubits idle (see Table~\ref{SI:TableLayerGates}
for detailed information). The qubits that are not selected for the
layer definition are left idle as well. The two-qubit Rzz gates are
natively supported~\cite{IBMFractionalGates} and have a duration of
88ns. All single-qubit Pauli rotations, including the Z rotation, are
decomposed as Rz-Sx-Rz-Sx-Rz sequences. The Sx gates have a uniform
duration of 32ns. The Rz gates are implemented
virtually~\cite{MCK2017WSCa}. Successive layer instances in the
circuit are separated by barriers and gates in each layer are
scheduled on an as-early-as-possible basis. As seen from the above
definition, the pulse schedule used to implement the gates in each
layer is not time independent. The Lindbladian that we learn therefore
models the overall layer operation instead of the individual gate
operations.

\begin{table}[ht]
\centering
\setlength{\tabcolsep}{5pt}
\begin{tabular}{|lll|ll|l|}
\hline
\multicolumn{3}{|l|}{Single-qubit gates} &
\multicolumn{2}{l|}{Two-qubit gates Rzz$(q_1, q_2, \theta)$} &
\multicolumn{1}{l|}{Idle qubits}\\
\hline
Ry(2, 0.24)&Rz(60, 0.38)&Ry(122, 0.18)&(0, 1, 0.32)&(69, 70, 0.12)&6\\
Ry(7, 0.22)&Rx(64, 0.36)&Ry(124, 0.28)&(3, 16, 0.16)&(72, 73, 0.30)&13\\
Rx(8, 0.12)&Rx(71, 0.30)&Ry(125, 0.34)&(4, 5, 0.24)&(78, 89, 0.36)&46\\
Rz(10, 0.12)&Rz(74, 0.12)&Ry(126, 0.30)&(14, 15, 0.26)&(80, 81, 0.20)&48\\
Rz(11, 0.20)&Rz(77, 0.32)&Rx(130, 0.16)&(17, 27, 0.34)&(93, 94, 0.38)&56\\
Rz(12, 0.16)&Ry(79, 0.38)&Ry(131, 0.24)&(19, 35, 0.36)&(98, 111, 0.14)&85\\
Rx(18, 0.32)&Rz(82, 0.34)&Rx(134, 0.10)&(20, 21, 0.38)&(99, 115, 0.14)&88\\
Rz(22, 0.24)&Rz(84, 0.30)&Rx(135, 0.34)&(25, 26, 0.14)&(100, 101, 0.10)&90\\
Rx(24, 0.38)&Rx(86, 0.18)&Rx(143, 0.24)&(29, 30, 0.12)&(103, 104, 0.10)&97\\
Rz(28, 0.12)&Rz(87, 0.40)&Ry(146, 0.30)&(32, 33, 0.28)&(105, 106, 0.32)&107\\
Rz(34, 0.14)&Rx(91, 0.22)&Rx(150, 0.36)&(38, 49, 0.20)&(113, 114, 0.26)&108\\
Rz(37, 0.30)&Rz(92, 0.26)&Ry(155, 0.38)&(40, 41, 0.28)&(123, 136, 0.24)&112\\
Ry(39, 0.16)&Ry(95, 0.26)& &(44, 45, 0.30)&(127, 128, 0.28)&118\\
Ry(42, 0.24)&Rz(96, 0.22)& &(47, 57, 0.30)&(132, 133, 0.38)&119\\
Ry(43, 0.16)&Rz(102, 0.36)& &(54, 55, 0.38)&(137, 147, 0.26)&129\\
Rz(50, 0.30)&Ry(109, 0.26)& &(59, 75, 0.14)&(140, 141, 0.14)&138\\
Ry(51, 0.28)&Rx(110, 0.10)& &(61, 76, 0.30)&(144, 145, 0.18)&139\\
Rz(52, 0.36)&Rx(116, 0.20)& &(62, 63, 0.24)&(148, 149, 0.16)&142\\
Ry(53, 0.22)&Rz(120, 0.24)& &(65, 66, 0.22)&(151, 152, 0.32)& \\
Ry(58, 0.20)&Rx(121, 0.16)& &(67, 68, 0.24)&(153, 154, 0.24)& \\
\hline
\end{tabular}

\caption{Gates and idle qubits defining the layer used to evaluate
  the learning protocol.}\label{SI:TableLayerGates}
\end{table}

\subsection{Readout error mitigation}\label{SI:TREX}

For learning, we start each circuit with single-qubit basis change
gates and then add the desired number of layer instances. We then
append single-qubit measurement-basis gates followed by twirled
measurements~\cite{BER2022MTa}. Twirling is implemented by sampling
various circuit instances of each target circuit and randomly applying
X or I gates prior to the measurement and applying the corresponding
classical operation on the measurement outcome. This helps diagonalize
the noise, which means that each Pauli-Z observable now has an
individual measurement fidelity. We estimate this fidelity based on
shot counts for circuits that only contain twirled
measurements. Mitigation is then done by dividing the measured Pauli-Z
expectation value by this readout fidelity. Overall, this method
scales well with growing numbers of qubits and, in the absence of
state-preparation errors, completely mitigates readout-errors in the
infinite shot limit for calibration. In the presence of
state-preparation error, the readout fidelity estimates also
incorporates state-preparation fidelities, which introduces errors in
the mitigated observable values.

\subsection{Fine-tuning protocol}\label{SI:FineTuning}

\begin{figure}[ht]
\centering
\begin{tabular}{cc}
% experiments/paper_finetuning.py
\includegraphics[width=0.45\textwidth]{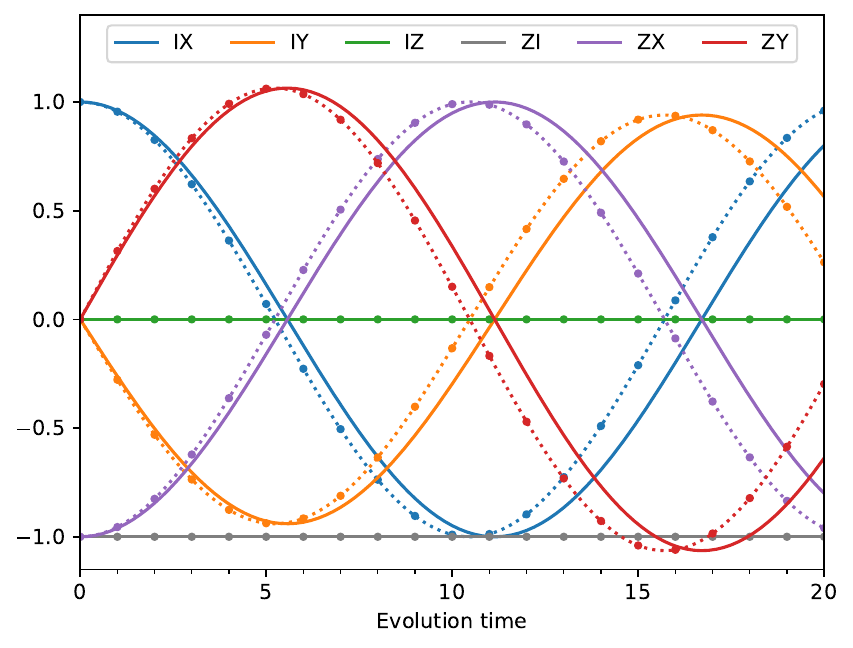}&
\includegraphics[width=0.45\textwidth]{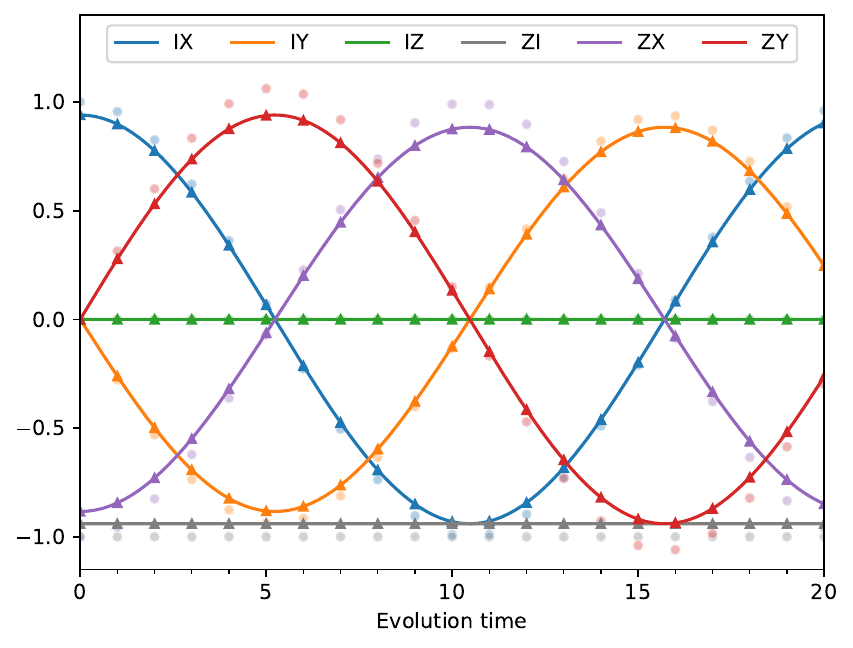}\\
({\bf{a}}) & ({\bf{b}})
\end{tabular}
\caption{Effects of model fine tuning for a setting with 3\%
  state-preparation error and no readout error. (a) Simulated
  observable values (markers) for the initial $\ket{1+}$ state
  following readout-error mitigation that is calibrated in the same
  setting, along with curve fits (dashed) and time evolution of the
  model (solid) learned from data with nine initial states. The ideal
  Lindbladian has a single ZZ Hamiltonian term with weight 0.15 and no
  dissipative terms. All simulation results assume infinite shots; (b)
  Time evolution of the fine-tuned model (solid) using unmitigated
  observable data (triangles). The readout-error mitigated data points
  (faint circles) are added for reference.}\label{SI:FigFineTuning}
\end{figure}

One way to check whether the learned noise model captures the physical
operation is to (locally) time evolve the model and check whether the
observable expectation values match the ones measured in the
experiment. Doing so in our experimental setting shows noticeable
differences. In part, these are explained by the fact that we would
need to know SPAM errors to accurately simulate the data based on the
learned model. Other errors are due to the effect of state-preparation
errors on readout-error mitigation. In order to better understand
this, we simulate learning of a simple two-qubit Lindbladian
corresponding to an Rzz(0.3) gate. For learning, we apply a 3\%
state-preparation error and no readout error, but nevertheless apply
readout-error mitigation calibrated with infinite shots in this
setting. The measured data points, fitted curve, and time-evolved
model are shown in Fig.~\ref{SI:FigFineTuning}a. The learned
Hamiltonian model coefficients are all zero, except for a single ZZ
term with value $0.141$, which is within a 7\% error compared to the
expected value of $\theta/2=0.15$, but results in noticeable misfits
to the simulated data.

Since the time-evolved model should ideally match the data (up to shot
noise) we fine tune the model parameters to reduce the misfit. Doing
so requires simulating time-evolution of the density matrix and can
therefore only be done for small (sub)sets of qubits. To enable fine
tuning we first introduce additional model parameters that capture the
state-preparation and measurement errors of the individual
qubits. Next, we implement a routine that evaluates the misfit, the
sum of squared residuals to the {\emph{unmitigated}} data, with
respect to the model parameters. Based on this, we evaluate the
gradient using forward finite differencing for each of the model
parameters with a step size $\delta=10^{-7}$. We then use
L-BFGS-B~\cite{BYR1995LNZa,ZHU1997BLNa} to minimize the misfit subject
to $[0,1]$ bound constraints on the state-preparation and measurement
errors, warm-stated with the learned model parameters. The results
based on the fine-tuned model, along with the unmitigated data are
shown in Fig.~\ref{SI:FigFineTuning}b. The updated Hamiltonian now
deviates from the ideal one with term errors of the order $10^{-7}$
and below. The time-evolved model now clearly matches the
data. Additional constraints, such as positive-semidefinite
constraints on $\beta$ could be added to ensure that the model remains
physical.

\subsection{Learned Hamiltonian and dissipative terms}\label{SI:LearnedTerms}

\begin{figure*}[!ht]
\centering
\includegraphics[width=0.95\textwidth]{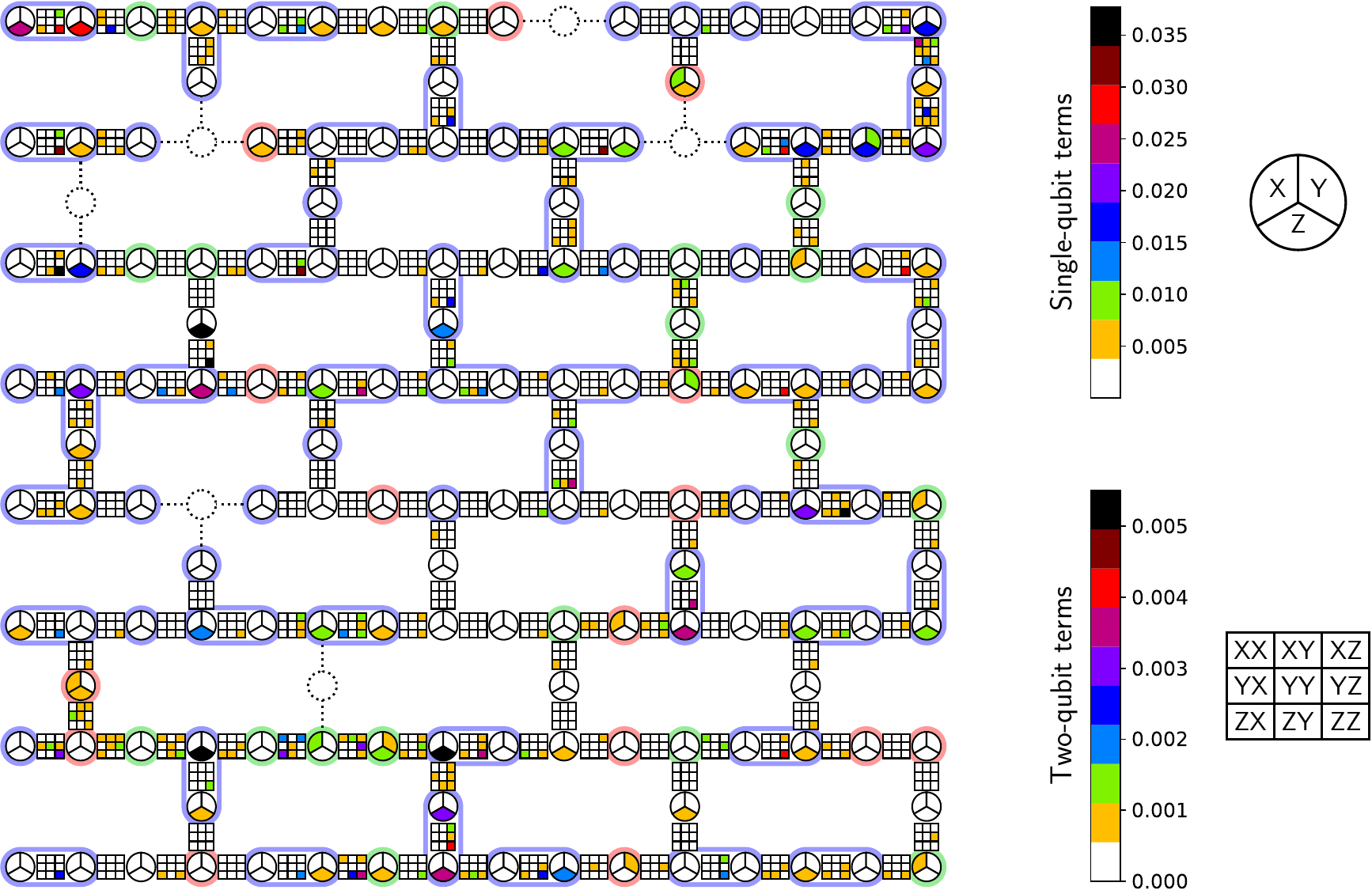}
\caption{Absolute error between the ideal Hamiltonian terms and the
  learned Hamiltonian terms following the qubit topology (circles for
  single-qubit terms, and a grid for all nine two-qubit Pauli terms on
  pairs of neighboring qubits). For reference we also display the
  single-qubit $R_x$, $R_y$, and $R_z$ gates (light red, green, and
  blue, respectively), and the two-qubit $R_{zz}$ gates (light
  blue).}\label{Fig:LearnedHamiltonianTermErrors}
\end{figure*}

The main text provides aggregated results of the 150-qubit Lindblad
model learned on the 156-qubit superconducting quantum processor
{\it{ibm\_pittsburgh}}. To provide better insight in the learned model
we consider for each term in the model, the absolute error between the
ideal (noiseless) circuit, and the values obtained by the learning
protocol.  The errors in the Hamiltonian terms, shown in
Figure~\ref{Fig:LearnedHamiltonianTermErrors}, capture the coherent
errors in the implementation of the gates. The largest errors in the
single-qubit Hamiltonian terms are roughly an order of magnitude
higher than those of the two-qubit terms. There is no obvious pattern
between the gate type and the errors in the Hamiltonian
terms. Overall, errors in the single-qubit Pauli Z terms do seem more
prevalent compared to the X and Y terms.

\begin{figure*}[!th]
\centering
\includegraphics[width=0.85\textwidth]{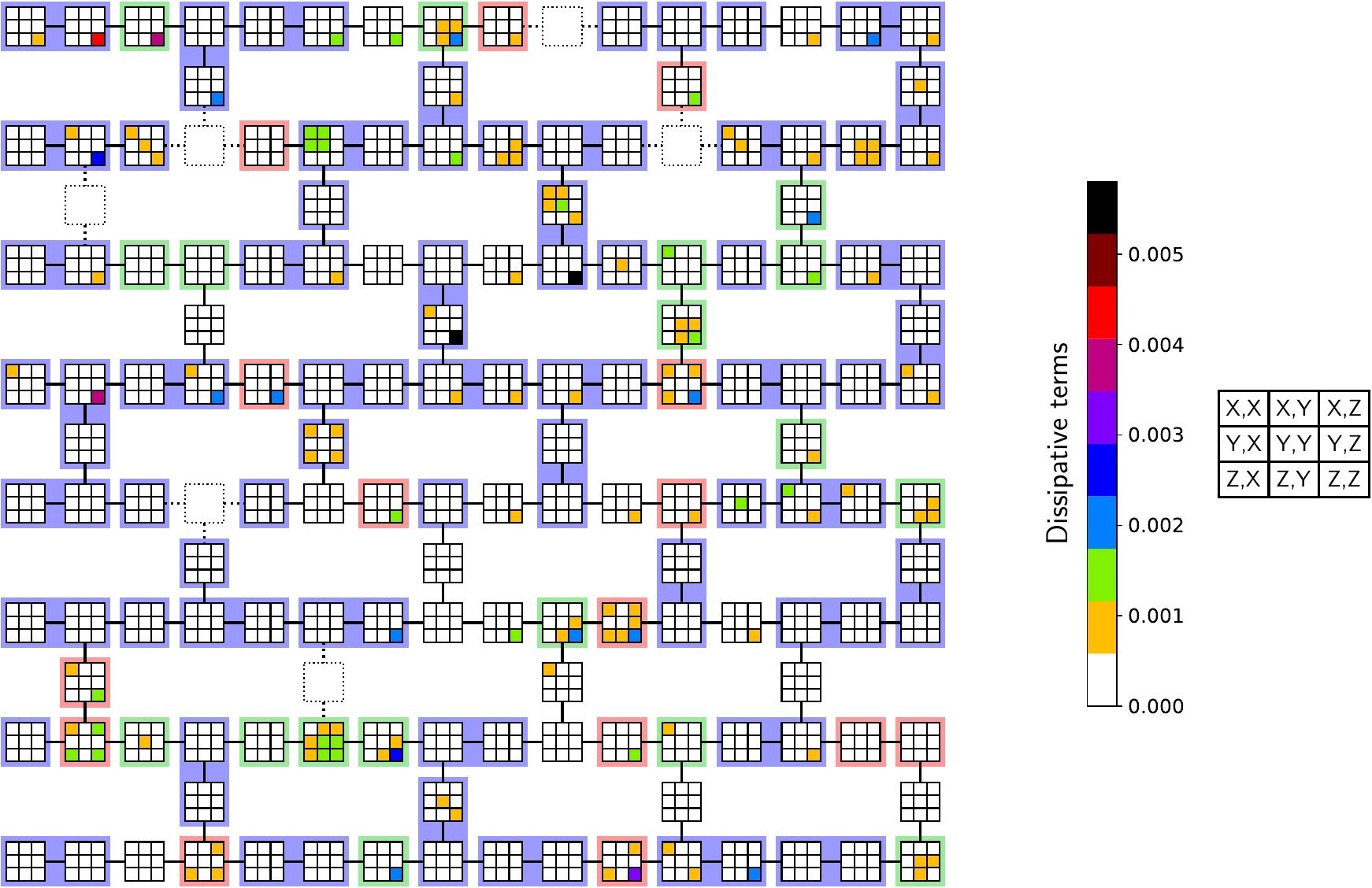}
\caption{Magnitude of the single-qubit dissipative terms plotted on
  the qubit topology. Each grid represents the entries in the block in
  the $\beta$ matrix associated with the given qubit. The mapping of
  the terms is shown in the grid on the right-hand side of the
  plot. The term values are normalized such that the layer of gates
  has a unit evolution time.}\label{Fig:LearnedDissipativeTermErrors}
\end{figure*}

Figure~\ref{Fig:LearnedDissipativeTermErrors} plots the absolute value
of the dissipative terms (for an ideal quantum circuit, the ideal
values should be zero). The per-qubit $\mathcal{D}_{Z,Z}$ terms,
equivalent to qubit dephasing, seem more pronounced. However, more
shots would be needed to faithfully reveal the structure of such
weaker dissipative terms.

\section{Experimentally learned error rates}\label{SI:Experiment_T1_T2}

\begin{figure}[!th]
\centering
\begin{tabular}{cc}
% experiments/paper5_experiments.py
\includegraphics[width=0.485\textwidth]{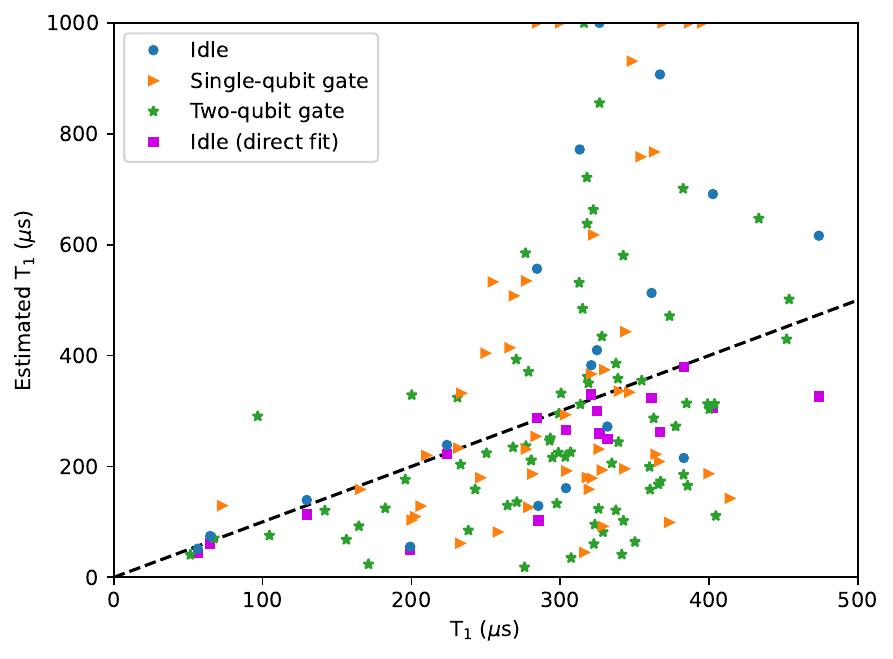}&
% experiments/paper5_experiments.py
\includegraphics[width=0.485\textwidth]{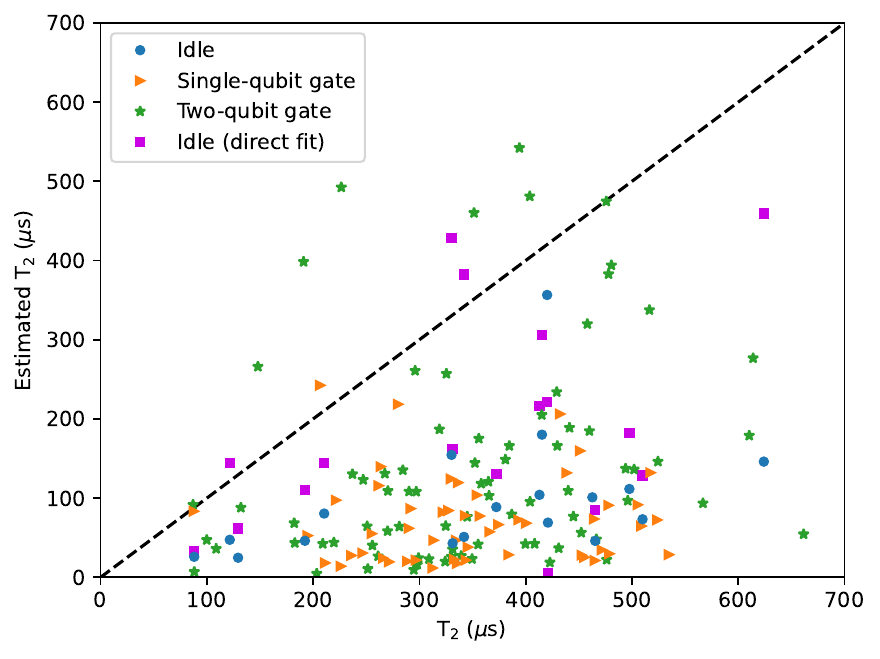}\\
({\bf{a}}) & ({\bf{b}})
\end{tabular}
\caption{Estimation of coherence times from the learned Lindblad
  model. (a) Estimated $T_1$ times against the values reported for
  {\it{ibm\_pittsburgh}} at the time of the experiment. The different
  markers correspond to the $T_1$ values obtained for idle qubits, as
  well as for qubits acted upon by a single-qubit or a two-qubit
  gate. The purple squares indicate the estimated $T_1$ value based on
  exponential curve fits to the time-series data on idle qubits with
  initial state $\ket{1}$ and measured in the computational basis. (b)
  $T_{2}$ times extracted from the learned Lindblad model plotted
  against the reported $T_2$ times. The purple squares are based on
  fits of idle qubits with initial state $\ket{+}$ and measured in the
  Pauli-X basis.}\label{SI:FigExperiment_T1_T2}
\end{figure}

We can analyze the terms in the learned Lindblad model over all 150
selected qubits to extract estimates for the $T_1$ and $T_{2\phi}$
coherence times. Specifically, we can use Eq.~\eqref{Eq:BetaBlock} to
extract the estimated $\beta_{\downarrow j}$ and $\beta_{\phi j}$
values from the block corresponding to qubit $j$ from the $\beta$
matrix, which is learned subject to positive-semidefinite
constraints. Following Section~\ref{SI:SyntheticSetup}, we then
estimate $T_1 = \tau / \beta_{\downarrow}$ with unit evolution time
$\tau = 88$ns and $T_2$ using
\[
\frac{1}{T_2} = \frac{1}{2T_1} + \frac{1}{T_{2\phi}},
\]
where $T_{2\phi} = \tau/ {\beta_{\phi}}$. The results shown in
Figure~\ref{SI:FigExperiment_T1_T2} plot the estimated values against
the $T_1$ and $T_2$ values reported for the {\it{ibm\_pittsburgh}}
device at the time of the experiment. In some cases, the
$\beta_{\downarrow}$ parameters assume values close to zero, which
results in large estimated relaxation times. For plotting purposes we
clip the estimated $T_1$ values at 1ms.

We can additionally use the data acquired for the idle qubits
initialized to $\ket{1}$ and measured in the computational basis to
directly estimate $T_1$. Likewise, we can estimate $T_2$ for idle
qubits initialized to $\ket{+}$ and measured in the Pauli-X basis. In
the figure, these estimates are indicated by the purple squares. More
accurate determination of these properties may require larger learning
depths. Alternatively, the parameters corresponding to $T_1$ and
$T_{2\phi}$ in the learned model could be replaced by values obtained
through direct measurements.

\begin{figure}[!hb]
\centering
% experiments/paper5_experiments.py
\includegraphics[width=0.6\textwidth]{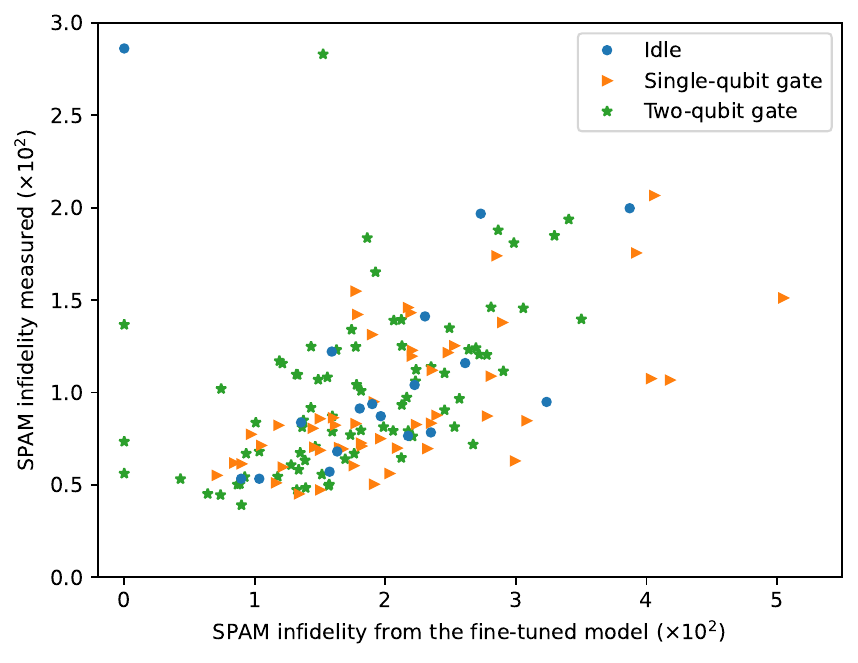}
\caption{Combined state-preparation and measurement infidelity for all
  model qubits. Estimated infidelity based on zero-depth
  readout-twirled circuits is plotted against the estimates from locally
  fine-tuned Lindblad models for individual qubits (idle qubits and
  qubits with a single-qubit gate) and for pairs of qubits (two-qubit
  gates).}\label{SI:FigExperiment_SPAM}
\end{figure}

As discussed in the main text, we can fine tune local Lindblad models
by augmenting them with parameters for state-preparation and
measurement errors and then applying local minimization of the
$\ell_2$ distance between the time-evolved model and the measured
data. In addition to obtaining a model that better fits the measured
data, we also obtain separate estimates for the state-preparation and
measurement error rates. Pauli-Z observable values measured for
zero-depth circuits prepared in the $\ket{0}$ or $\ket{1}$ state with
readout twirling, provide information about the overall SPAM fidelity,
which is the product of the state-preparation fidelity and the
measurement fidelity. In Fig.~\ref{SI:FigExperiment_SPAM} we plot the
SPAM infidelity (one minus the fidelity) of the zero-depth estimates
against those obtained from the fine-tuned local models. Note that we
did not incorporate the measured SPAM fidelities during fine tuning.

\clearpage

\twocolumngrid
\bibliographystyle{unsrt}
\bibliography{bibliography}

\end{document}